\documentclass{article}
\PassOptionsToPackage{numbers,compress}{natbib}
\usepackage[main,final]{neurips_2026}
\usepackage{natbib}

\usepackage[utf8]{inputenc} 
\usepackage[T1]{fontenc}   
\usepackage{hyperref}
\usepackage{url}
\usepackage{booktabs}
\usepackage{amsfonts}
\usepackage{nicefrac}
\usepackage{microtype}
\usepackage[table,dvipsnames]{xcolor}
\usepackage{graphicx}
\usepackage{soul}
\usepackage{amsmath}
\usepackage{wrapfig}
\usepackage{enumitem}

\usepackage{caption}

\usepackage{float}
\usepackage[normalem]{ulem}

\hypersetup{
	colorlinks=true,
	linkcolor=black,
	citecolor=MidnightBlue,
	urlcolor=blue,
}

\usepackage[nameinlink]{cleveref}
\Crefname{appsec}{Appendix}{Appendices}

\definecolor{StrongRow}{RGB}{202,244,214} 
\definecolor{MediumRow}{RGB}{238,248,255}
\definecolor{WeakRow}{RGB}{252,226,226}

\title{Capturing Membrane Dynamics and Spike Timing of Human Neurons at Scale using Neural Operators
}

\author{%
  Luca Ghafourpour$^{1}$\thanks{These authors contributed equally to this work.} \hspace{3mm}
  Cindy Zhao$^{2}$\footnotemark[1] \hspace{3mm}
  Valentin Duruisseaux$^2$ \hspace{3mm}
  Bahareh Tolooshams$^{3,4}$ \\
  \vspace{1pt}
  \textbf{Philip H. Wong}$^5$ \hspace{3mm}
  \textbf{Costas A. Anastassiou}$^{2, 5,6}$ \hspace{3mm}
  \textbf{Anima Anandkumar}$^2$ \vspace{4pt}\\
  $^1$University of Cambridge\quad
  $^2$California Institute of Technology\quad
  $^3$University of Alberta\\
  $^4$Alberta Machine Intelligence Institute (Amii)\quad
  $^5$Cedars-Sinai Medical Center\\
  $^6$Archimedes AI, Athena Research Center
}

\makeatletter
\newcommand{\sublabel}[2]{\begingroup\protected@edef\@currentlabel{\thetable#2}\protected@edef\cref@currentlabel{[table][\arabic{table}][]\thetable#2}\label{#1}\endgroup}
\makeatother

\makeatletter
\renewcommand{\section}{\@startsection{section}{1}{\z@}{-1.6ex \@plus -0.4ex \@minus -0.2ex}{1.0ex \@plus 0.2ex \@minus 0.2ex}{\large\bf\raggedright}}
\renewcommand{\subsection}{\@startsection{subsection}{2}{\z@}{-1.4ex \@plus -0.4ex \@minus -0.2ex}{0.5ex \@plus 0.2ex}{\normalsize\bf\raggedright}}
\renewcommand{\paragraph}{\@startsection{paragraph}{4}{\z@}{1.0ex \@plus 0.4ex \@minus 0.2ex}{-1em}{\normalsize\bf}}
\makeatother

\begin{document}
\maketitle

\begin{abstract}
Characterizing the properties of neuronal cell types is central to understanding brain function. Yet, biophysically realistic neuron models remain difficult to scale to large populations while preserving the richness of cellular dynamics and within-cell-type variability. We present a cell-type-specific neural-operator framework for scalable neuronal population modeling. Rather than training separate surrogate models for individual neurons, the proposed framework learns a shared operator across multiple modeled neurons from the same cell type, conditioned on biologically interpretable electrophysiological features. Relative to the original single-neuron \texttt{NOBLE} setting, learning across heterogeneous neurons requires a richer electrophysiological conditioning representation and a training objective designed to preserve diverse spiking dynamics. This enables efficient generation of large ensembles of somatic voltage responses while preserving variability across cells and model configurations. We evaluate the approach using membrane dynamics metrics, action potential waveform metrics, spike timing accuracy, input-output firing relationships, and electrophysiological feature distributions. Our framework reproduces subthreshold membrane dynamics, spike waveforms, and firing-rate responses across major cortical inhibitory neuron types, while accurately preserving spike timing despite the sensitivity of threshold-crossing dynamics. The predicted electrophysiological feature distributions show strong agreement with detailed simulator outputs and human experimental recordings, including high explained variability in properties such as inter-spike intervals, firing frequency, action-potential shape, and post-spike recovery. By enabling fast ensemble simulation of cell-type-specific neuronal responses, the proposed framework provides a scalable tool for studying cellular diversity, probing variability near functional thresholds, and generating synthetic but biophysically grounded neuronal populations.
\end{abstract}

\vspace{3mm}

\section{Introduction}\label{sec:introduction}

\looseness=-1 Understanding how neural computation arises from the diversity of neuronal cell types remains a fundamental challenge in neuroscience~\citep{douglas1995recurrent,buzsaki2010neural,luo2021architectures,siletti2023transcriptomic,zhang2025cell}. Neural circuits comprise hundreds of distinct cell types with diverse gene-expression, morphological, and electrophysiological profiles~\citep{sunkin2012allen,gouwens2020integrated}, and human neurons in particular exhibit substantial electrophysiological variability. Even identical current inputs to the same neuron can produce markedly different burstiness, adaptation, and spike-rate statistics~\citep{berg2021human,mainen1995reliability,mainen1996influence,anastassiou2011ephaptic,lee2024cell}. Multimodal datasets now enable increasingly detailed characterization of neuronal cell types~\citep{sunkin2012allen,gouwens2020integrated,berg2021human,chartrand2023morphoelectric,lee2023signature}, but linking this diversity to computation remains poorly understood, especially in the human brain, where experimental access and monitoring capabilities are limited. The challenge is amplified for cell types defined jointly across transcriptomic, morphological, and electrophysiological modalities, whose gene-expression signatures co-segregate with biophysical and morphological properties, and for which population-level variability makes generalization to unseen stimuli and model configurations especially difficult. How neuronal cell types organize, generate canonical computations, and contribute to neurological disorders therefore remains an open problem, and accurate single-neuron and cell-type representations are essential for developing circuit-level models of the brain.

\looseness=-1 Computational neuroscience aims to bridge this gap by linking multimodal cellular properties to functional dynamics. Biorealistic multi-compartment models integrate morphology, ion channel distributions, and electrophysiological recordings, enabling controlled perturbations that are often experimentally impossible~\citep{reimann2013biophysically,buchin2022multi,nandi2022single}. These conductance-based models are deterministic, so variability is typically approximated by fitting multiple hall-of-fame (HoF) models per experiment, each reproducing a subset of experimental features~\citep{buchin2022multi,nandi2022single,druckmann2007novel,van2016bluepyopt,mosher2020cellular}. Obtaining HoFs can require hundreds of thousands of CPU hours, and interpolating between them is unreliable because the dynamics of these biophysical partial differential equation (PDE) models are highly sensitive to their parameters~\citep{buchin2022multi,nandi2022single,achard2006complex}. Existing approaches therefore remain limited in their ability to scalably and realistically capture neuronal variability.

\looseness=-1 Neural operator surrogates offer an approach to overcoming these challenges. \texttt{NOBLE}~\citep{ghafourpour2025noble} conditions a neural operator on current injections and interpretable neuron-model features to predict somatic voltage responses across the HoF models of a single neuron, with a speedup of several orders of magnitude over the numerical solver (\Cref{sec:related_work}). However, extending this approach to biologically defined neuronal populations requires capturing both HoF-dependent response variability and structured heterogeneity across neurons within the same transcriptomic cell type. The original two-dimensional conditioning on the threshold current $I_{\mathrm{thr}}$ and local frequency-current (F-I) slope $s_{\mathrm{thr}}$, though sufficient within one neuron, becomes ambiguous at this level, since HoF models from different neurons can share these coordinates while exhibiting different dynamics (\Cref{sec:feature_selection}).

\begin{figure}[t]
    \vspace{-4mm}
    \centering
    \includegraphics[width=0.90\linewidth]{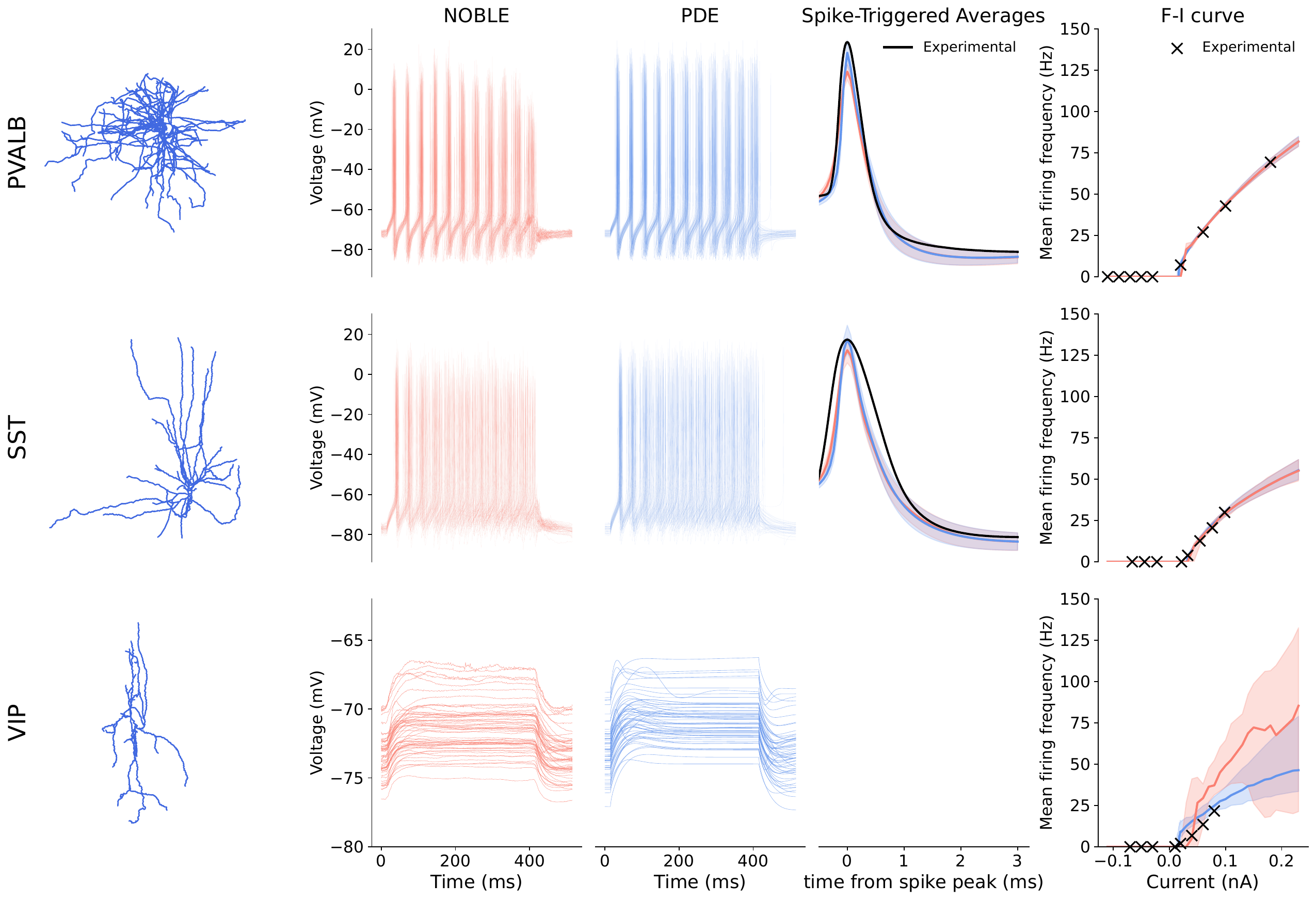}
    \caption{\textbf{Performance across transcriptomic types.} From left to right, columns show: (1) cell morphology; (2) voltage traces predicted by \texttt{NOBLE}; (3) voltage traces generated by PDE simulations; (4) spike-triggered voltage averages from \texttt{NOBLE}, PDE simulations, and experimental recordings; and (5) F-I curves from \texttt{NOBLE}, PDE simulations, and experimental recordings.
        For columns (4) and (5), the solid line represents the median, and the shaded area represents the interquartile range of the ensemble.}
    \label{fig:performance_overview}
    \vspace{-2mm}
\end{figure}

\textbf{Contributions.} Using a rich human electrophysiological dataset spanning major cortical inhibitory populations, we learn a single surrogate shared across all modeled neurons of a transcriptomic cell type:
\begin{itemize}[leftmargin=*,topsep=1pt,itemsep=1pt,parsep=0pt]
    \item \textbf{A cell-type-level biologically informed embedding.} We select compact, interpretable features that capture within-cell-type variability while providing complementary information. Reverting to the original two-feature representation is found to increase the inverse interspike interval (ISI) error nearly threefold (\Cref{tab:embedding_comparison}).
    \item \textbf{A spike-weighted training objective.} A sigmoid-gated relative $\mathrm{L2}$ loss emphasizes spike-region errors, improving spike timing and action-potential dynamics without sacrificing subthreshold accuracy (\Cref{sec:loss}).
    \item \textbf{Population-level validation against simulations and human recordings.} Across three classes of inhibitory neurons, waveforms and feature distributions match PDE-based HoF simulations with smaller error than the PDE-to-experiment discrepancy, while inference is up to $3{,}640\times$ faster than the numerical solver (\Cref{fig:performance_overview}, \Cref{sec:results}).
\end{itemize}

\section{Related Work}\label{sec:related_work}

\looseness=-1 \textbf{Biophysical neuron models.} Evolutionary multi-objective optimization~\citep{miettinen1999nonlinear,druckmann2007novel,van2016bluepyopt} and, more recently, gradient-based methods~\citep{deistler2025jaxley} fit detailed multi-compartment models to multimodal recordings~\citep{buchin2022multi,nandi2022single}. Each fitted model is deterministic and costly to simulate, so ensembles of HoF models approximate the trial-to-trial variability observed experimentally.

\looseness=-1 \textbf{Machine learning surrogates.} The computational cost of numerical integration and evolutionary optimization has motivated machine learning surrogates for single-cell electrophysiology, primarily in simplified point-neuron models such as the FitzHugh--Nagumo~\citep{fitzhugh1955mathematical,nagumo1962active} and Hodgkin--Huxley~\citep{hodgkin1952quantitative}. Conventional and physics-informed neural networks~\citep{Raissi2019,Raissi2017Part1,Raissi2017Part2} reproduce these dynamics but remain largely specific to deterministic formulations, often require retraining for new stimuli or model configurations~\citep{su2021deep,rudi2022parameter,ferrante2022physically,kuptsov2023discovering,werneck2023replacing,pandey2024efficient}, and have not been shown to capture the full range of electrophysiological behavior, especially in the nonlinear firing threshold regime. Beyond these neuroscience-specific limitations, standard neural networks also face a more fundamental representational limitation: they learn mappings between finite-dimensional vectors, which can limit their ability to model physical systems more naturally described by functions and relationships between functions~\citep{azizzadenesheli2024neural}. As a result, they may overfit to the training discretization and generalize poorly out of distribution.

\looseness=-1 \textbf{Neural operators.} These limitations motivate neural operators~\citep{azizzadenesheli2024neural,kovachki2023neural,Berner2025Principles}, which learn mappings between function spaces with universal approximation guarantees~\citep{Kovachki2021_2}. In neuroscience, they have been applied to Hodgkin--Huxley dynamics~\citep{centofanti2024learning} without trial-to-trial or biological variability. \citet{ghafourpour2025noble} introduced \texttt{NOBLE}, a neural operator with biologically informed latent embeddings that predicts somatic voltage responses across the HoF models of a neuron without requiring a separate surrogate for each model. \texttt{NOBLE} captured subthreshold and spiking dynamics, generalized to unseen HoF models, supported latent-space interpolation for generating novel ensembles, reproduced electrophysiological features within the variability observed in human cortex recordings, and achieved speedups of several orders of magnitude over the numerical solver. We build on this backbone and extend it from one neuron to an entire transcriptomic cell type.

\section{PDE-Based Modeling of Neuronal Dynamics}\label{sec:background}

\looseness=-1 We use a Patch-seq dataset of human cortical inhibitory neurons from the middle temporal gyrus~\cite{lee2023signature}, focusing on the three transcriptomic classes for which biophysical models were available: parvalbumin (PVALB), somatostatin (SST), and vasoactive intestinal peptide (VIP). Each of these cell types exhibits distinct cortical dynamics. PVALB neurons are specialized fast-spiking inhibitory interneurons that orchestrate high-frequency gamma oscillations (30--80 Hz)~\cite{kim2015cortically}. SST neurons, while also capable of supporting gamma oscillations, are generally associated with slower beta and theta oscillations~\cite{jackson2024amygdala,park2020optogenetic}. VIP neurons are key mediators in disinhibitory circuit motifs, where they selectively target SST neurons to release excitatory cells from dendritic inhibition~\cite{pi2013cortical,pronneke2015characterizing,tremblay2016gabaergic,veit2023cortical}.

\looseness=-1 For each neuron, we construct biophysically detailed, PDE-based ``all-active'' models grounded in cable theory~\cite{koch2004biophysics,buchin2022multi,nandi2022single} from high-resolution morphological reconstructions and electrophysiological recordings~\cite{gouwens2020integrated,berg2021human,lee2023signature}, simulated with BioNet~\cite{gratiy2018bionet} and NEURON~\cite{hines2001neuron}. Multi-objective evolutionary optimization~\cite{nandi2022single} generated 60 HoF models per neuron whose ion-conductance parameters replicate the electrophysiological features of interest. Further details on the dataset and optimization protocol are given in \Cref{app:hof_models}, and the features are defined in \Cref{app:e_features_appx}.

\section{Method}\label{sec:method}

\subsection{The \texttt{NOBLE} Backbone}\label{sec:noble}

\begin{figure}[t]
    \vspace{-2mm}
    \centering
    \includegraphics[width=0.85\linewidth]{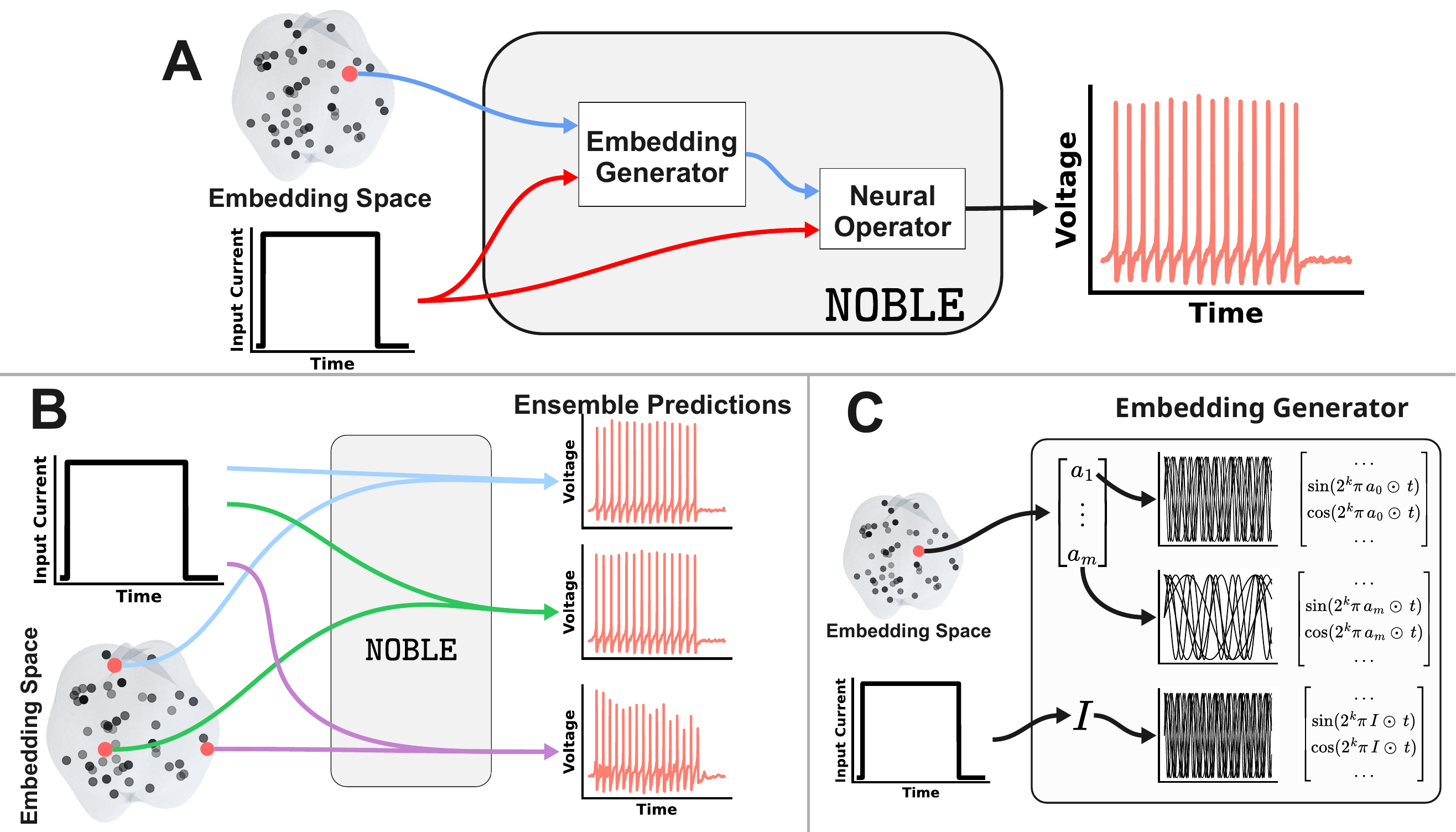}
    \caption{\texttt{NOBLE} (extracted from \citep{ghafourpour2025noble}).
        \textbf{A}) A current injection and neuron model features are embedded before being passed through a neural operator to produce a somatic voltage response prediction.
        \textbf{B}) \texttt{NOBLE} can be queried in parallel with different model latent representations to produce ensemble predictions.
        \textbf{C}) The embedding generator encodes specified neuron features and the input current as a stack of trigonometric time series, which are fed as input channels to the neural operator.}
    \label{fig:noble_architecture}
    \vspace{-1mm}
\end{figure}

\looseness=-1 \texttt{NOBLE}~\citep{ghafourpour2025noble} learns a direct mapping from an input current and a continuous, interpretable latent space of neuron features to the resulting somatic voltage trace. It uses a Fourier Neural Operator (FNO)~\citep{li2020fno,duruisseaux2025FNOGuide}, which is well suited to electrophysiological data because recordings and PDE simulations are sampled on equidistant time grids (architecture shown in \Cref{fig:fno_architecture}).

The latent representation in \texttt{NOBLE} is constructed from user-defined, interpretable neuron characteristics. \citet{ghafourpour2025noble} summarized cellular excitability using the F-I curve, which maps injected current to firing rate. Its threshold current $I_{\mathrm{thr}}$ and local threshold slope $s_{\mathrm{thr}}$ capture variability across HoF models, yielding interpretable latent coordinates $(I_{\mathrm{thr}}, s_{\mathrm{thr}})$ for HoFs. To make these low-dimensional features compatible with the FNO, \texttt{NOBLE} uses a frequency-modulated embedding. More precisely, a feature $p$ is encoded as a stack of trigonometric time series

\begin{equation}\label{eq:embedding}
    \gamma(p, t) =
    \left[
    \sin(2^0 \pi p \odot t), \
    \cos(2^0 \pi p \odot t), \
    \ldots, \
    \sin(2^{K-1} \pi p \odot t), \
    \cos(2^{K-1} \pi p \odot t)
    \right],
\end{equation}

for some integer $K>0$, where the frequencies are modulated by the feature $p$, $t$ denotes the discretized time coordinates, and $\odot$ indicates element-wise multiplication with appropriate broadcasting. The model features and current amplitude are encoded separately and stacked as input channels. This sinusoidal embedding is synergistic with FNOs, which operate in the frequency domain to learn mappings between functions, and acts as a form of spectral lifting, translating low-dimensional biological features into a richer frequency-domain representation that the neural operator can efficiently process. Our models receive $53$ input channels, namely the input current stimulus, its embedding with $K = 2$ frequencies, and the embedding of each feature with $K = 4$ frequencies.

\looseness=-1 \texttt{NOBLE} is well suited to this setting because it maps current injections and biologically meaningful latent features to somatic voltage responses in a unified, interpretable framework. A single operator can generate diverse, biophysically plausible responses to the same input current by varying the latent representation (\Cref{fig:noble_architecture}B), capturing both subthreshold and spiking dynamics without retraining. Its continuous latent space also supports interpolation between HoF models, enabling realistic responses beyond those found by evolutionary optimization. Thus, \texttt{NOBLE} provides a scalable surrogate for families of biophysical neuron models with latent coordinates that can be sampled, interpolated, and analyzed. In this work, we retain this operator backbone but modify the embedding representation and training objective to support a single shared model across heterogeneous neurons within a cell type.

\subsection{Feature Selection for Informed Model Embeddings}\label{sec:feature_selection}

\looseness=-1 In the single-neuron setting of~\citet{ghafourpour2025noble}, the \texttt{NOBLE} latent space was parameterized by two physically meaningful features, $I_{\mathrm{thr}}$ and $s_{\mathrm{thr}}$. This compact embedding was effective for the $\sim$60 HoF configurations of one cell, which were approximately uniformly distributed in the convex hull of their $(I_{\mathrm{thr}}, s_{\mathrm{thr}})$ representations. However, in our multi-neuron setting, the model must represent hundreds of HoF configurations per cell type (810 for PVALB), spanning much broader electrophysiological heterogeneity within a transcriptomic class. A two-dimensional embedding is therefore insufficient, since distinct HoF models can occupy nearby latent points despite markedly different somatic dynamics. We instead enrich the latent representation with additional electrophysiological features, producing a higher-dimensional manifold in which latent proximity better reflects similarity in neuronal dynamics. The purpose of this richer representation is not simply to increase the embedding dimensionality, but to resolve ambiguities that arise when a single operator is trained across heterogeneous neurons.
The resulting electrophysiological embeddings capture cell-to-cell variability across the full population within each cell type.

\looseness=-1 To construct this representation, we simulate each HoF model under two 1\,s DC current steps ($0.28$\,nA and $-0.11$\,nA, probing suprathreshold spiking and subthreshold passive responses) and extract the features listed in \Cref{app:e_features_appx}. We rank candidate features by their mean absolute deviation across all HoF models of the cell type, $\mathrm{MAD}(F) = \frac{1}{N} \sum_{i=1}^{N} |F_i - \bar{F}|$, and then sequentially retain a feature only if its absolute correlation with all previously selected features is below a fixed threshold. This follows the maximum-relevance minimum-redundancy principle, using within-cell-type dispersion as the relevance criterion. We use this as an interpretable heuristic rather than an optimality criterion. Per-feature scores, selections for all cell types, and further discussion of this heuristic are given in \Cref{app:embedding_analysis}. For PVALB, the heuristic selects \texttt{intercept}, \texttt{Spikecount}, \texttt{inv\_first\_ISI}, \texttt{AHP\_depth}, \texttt{mean\_AP\_amplitude}, and \texttt{voltage\_base}. The number of embedded features is treated as a hyperparameter and evaluated in \Cref{tab:embedding_ablation}.

\subsection{Spike-Weighted Training Objective}\label{sec:loss}

Action potentials occupy only a small fraction of each trace, so the relative $\mathrm{Lp}$ loss used by~\citet{ghafourpour2025noble} under-weights errors in spike timing and waveform shape. We add a sigmoid-gated term that places greater weight on the spiking region while preserving the original objective over the full trace:

\begin{equation}\label{eq:soft_threshold_lp_main}
    \mathcal{L}(x, y) = \text{Lp}(x, y) + w\, \frac{(\sum_i g(y_i)\, |x_i - y_i|^p)^{1/p}}{(\sum_i g(y_i)\, |y_i|^p)^{1/p} + \epsilon}, \qquad g(y_i)=\sigma\!\left(\frac{y_i-\tau}{s}\right),
\end{equation}

where $\tau=-0.02$ is the spike threshold, $s=0.05$ sets the sharpness of the gate, and we use $p=2$ and $w=10$. We selected this objective from a range of standardized, gated, and Fourier-augmented $\mathrm{Lp}$ losses via small-scale overfitting studies (\Cref{app:overfitting}), followed by full-scale training of the two strongest candidates (\Cref{sec:loss_results}).

\subsection{Training and Evaluation Setup}\label{sec:setup}

\looseness=-1 We train a separate model per cell type: PVALB (16 cells, 810 HoF models, 427{,}500 traces), VIP (14 cells, 770 HoF models, 373{,}500 traces), and SST (7 cells, 410 HoF models, 184{,}500 traces). Test sets consist of held-out randomly sampled DC stimuli applied to HoF models seen during training ($47{,}500$ PVALB, $41{,}500$ VIP, and $20{,}500$ SST samples). For out-of-distribution evaluation on PVALB, we additionally hold out ten HoF models (five from each of two cells) and two entire cells ($60$ HoF models each). We report relative $\mathrm{L2}$ voltage error, relative errors of electrophysiological features (\Cref{app:e_features_appx}), and missed-firing and missed-non-firing rates. Missed firing is the rate at which a predicted trace contains no spikes although its target fires, and missed non-firing is the rate at which it contains spikes although its target does not. Architecture and optimization details are given in \Cref{app:implementation}. We additionally benchmark computational efficiency under a fixed hardware configuration. Using a workstation equipped with a single NVIDIA RTX 4090 GPU with 24\,GB of VRAM, an AMD Ryzen 9 7900X CPU, and 64\,GB of system RAM, the numerical solver requires approximately $36{,}200$\,s to generate $10{,}000$ predictions, processing one prediction at a time.
For the same number of predictions, our model achieves an approximately $1{,}940\times$ speedup without batching.
With batched inference using a batch size of 5, the speedup increases to approximately $3{,}640\times$.

\section{Results}\label{sec:results}

\vspace{-2mm}

\subsection{Accuracy Across Cell Types and Relative to Experimental Variability}\label{sec:test_performance}

\begin{table}[t] \vspace{-2mm}
    \centering
    \caption{\textbf{(a)} Test-set performance of \texttt{NOBLE} across PVALB, SST, and VIP neurons.
        Values report relative $\mathrm{L2}$ errors in voltage traces and selected electrophysiological metrics on held-out responses.
        \textbf{(b)} Relative $\mathrm{L2}$ discrepancy between PDE simulations and experimental recordings across inhibitory neuron classes, evaluated using the same metrics as in a).}
    \vspace{1mm}
    \label{tab:error_budget}
    \sublabel{tab:test_performance}{a}
    \sublabel{tab:PDE_experiment}{b}
    \renewcommand{\arraystretch}{1.02}
    \setlength{\tabcolsep}{5pt}
    \resizebox{\linewidth}{!}{%
        \begin{tabular}{@{}lcccccccc@{}}
            \toprule
            \textbf{Cell type}
            & \textbf{Voltage}
            & \shortstack{\textbf{inv ISI}\\\textbf{values}}
            & \shortstack{\textbf{mean AP}\\\textbf{amplitude}}
            & \shortstack{\textbf{steady-state}\\\textbf{voltage}}
            & \shortstack{\textbf{AP}\\\textbf{width}}
            & \shortstack{\textbf{Spikecount}}
            & \shortstack{\textbf{Missed}\\\textbf{firing}}
            & \shortstack{\textbf{Missed}\\\textbf{non-firing}}\\
            \midrule
            \multicolumn{9}{@{}l}{\textit{(a) \texttt{NOBLE} vs.\ PDE}}\\
            PVALB & 1.73\% & 16.97\% & 8.73\%  & 0.25\% & 10.54\% & 19.27\% & 1.87\% & 0.00\% \\
            SST   & 1.85\% & 8.07\%  & 8.53\%  & 0.42\% & 11.08\% & 12.02\% & 2.36\% & 0.00\% \\
            VIP   & 2.69\% & 11.96\% & 12.75\% & 0.53\% & 13.43\% & 17.96\% & 2.60\% & 0.00\% \\
            \midrule
            \multicolumn{9}{@{}l}{\textit{(b) PDE vs.\ experiment}}\\
            PVALB & 12.97\% & 47.13\% & 16.38\% & 13.59\% & 33.97\% & 42.50\% & 0.59\% & 8.00\% \\
            SST   & 15.43\% & 21.10\% & 14.75\% & 15.95\% & 86.92\% & 34.81\% & 0.00\% & 2.81\% \\
            VIP   & 24.46\% & 51.39\% & 19.93\% & 25.18\% & 17.75\% & 49.52\% & 0.17\% & 2.37\% \\
            \bottomrule
        \end{tabular}}   \vspace{0.5mm}
\end{table}

\begin{figure}[t]
    \vspace{-2mm}
    \centering
    \includegraphics[width=0.85\linewidth]{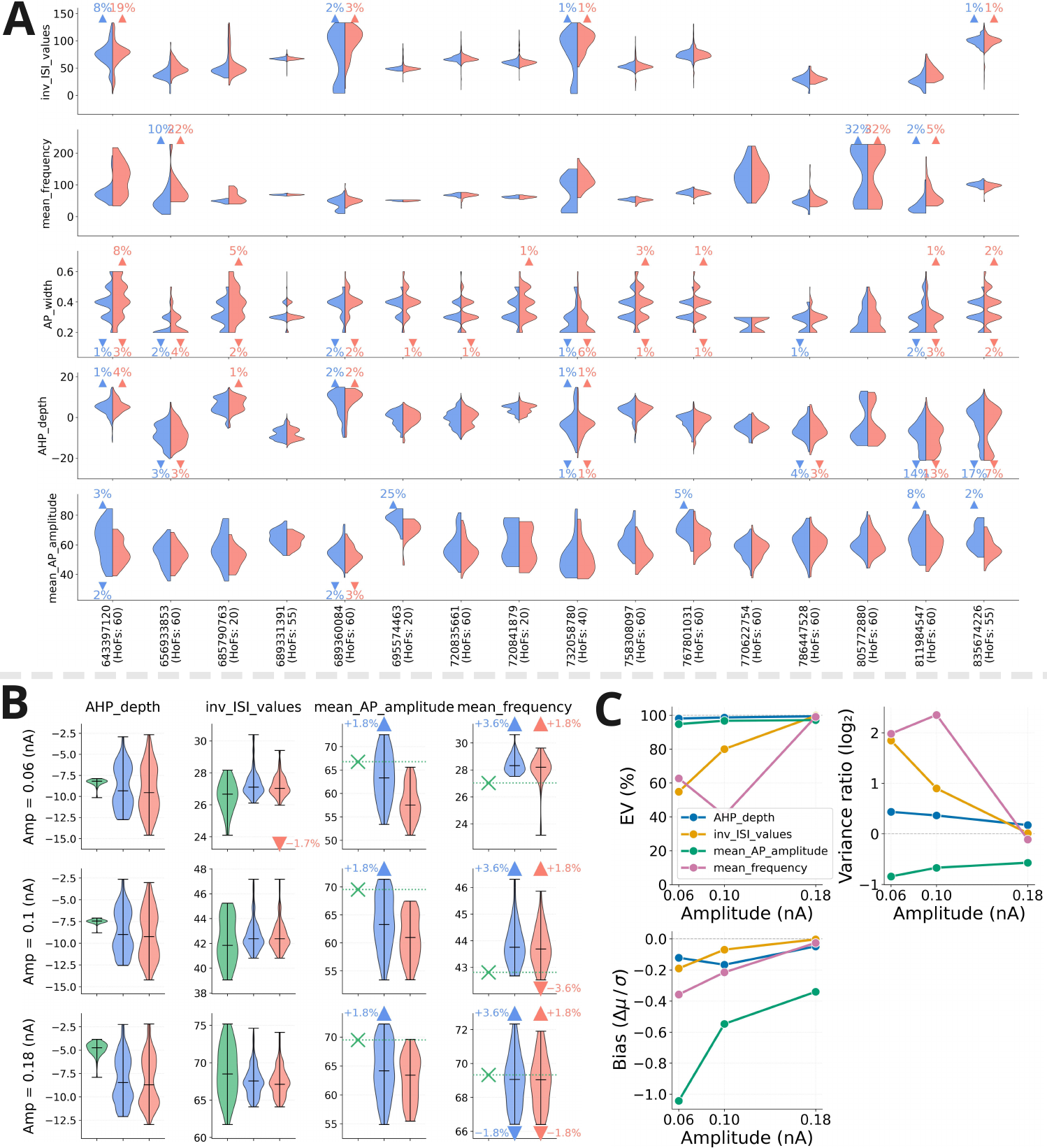}
    \caption{\textbf{A)} Distribution of key electrophysiology features for all PVALB neurons from \texttt{NOBLE} (red) and PDE simulations (blue).
        \textbf{B)} Distribution of key electrophysiology features for a representative PVALB neuron from \texttt{NOBLE}, PDE simulations, and experimental recordings (green) across three input amplitudes.
        Carets mark the fraction of values outside the displayed axis range.
        \textbf{C)} Explained variability between \texttt{NOBLE} and PDE distributions: explained variance (EV, \%), variance ratio ($\log_2 \mathrm{Var}(\texttt{NOBLE}) / \mathrm{Var}(\text{PDE})$), and normalized mean bias ($\Delta\mu / \sigma$).}
    \label{fig:efeature_distributions}
    \vspace{-2.5mm}
\end{figure}

\looseness=-1 We first evaluated whether the trained \texttt{NOBLE} models accurately reproduce held-out voltage responses. As shown in Table~\ref{tab:test_performance}, \texttt{NOBLE} achieves low test relative L2 voltage errors across PVALB, SST, and VIP neurons, while preserving spike timing, subthreshold voltage properties, action potential (AP) shape, and spike-count statistics across cell types. These results show that \texttt{NOBLE} accurately predicts responses to held-out input current amplitudes for HoF models represented during training. Generalization to unseen HoF parameterizations and entirely unseen cells is evaluated in \Cref{sec:analysis}.

\looseness=-1 \Cref{fig:performance_overview} provides a qualitative overview of this performance across PVALB, SST, and VIP neurons. Despite their distinct morphologies and response regimes, \texttt{NOBLE} closely reproduces the voltage traces generated by the PDE simulator, capturing both the sustained high-frequency firing of PVALB and SST neurons and the predominantly subthreshold and delayed depolarizing responses of VIP neurons. This suggests that \texttt{NOBLE} learns cell-type-specific response dynamics rather than relying on a single stereotyped voltage pattern. Predicted spike-triggered waveforms closely track the PDE-derived averages and remain consistent with experimental recordings, and the F-I curves reproduce both firing onset and gain. Together, these results show that \texttt{NOBLE} captures trace-level dynamics, spike waveform structure, and population-level input-output behavior across diverse inhibitory neuron classes.

\looseness=-1 To contextualize surrogate error, we compare the \texttt{NOBLE} errors in \Cref{tab:test_performance} with the PDE-to-experiment discrepancy in \Cref{tab:PDE_experiment} and with the experimental cell-to-cell and trial-to-trial variability in \Cref{tab:cell_to_cell,tab:trial_to_trial}. For all reported metrics except missed firing, \texttt{NOBLE} incurs lower error relative to its PDE targets than the discrepancy between the PDE simulations and experimental recordings. The PVALB missed-firing error of $1.87\%$ exceeds the PDE-to-experiment discrepancy of $0.59\%$, but remains below the cell-to-cell and trial-to-trial variability of $9.23\%$ and $20.00\%$. Surrogate error is therefore generally not the dominant source of mismatch with experiment, and improving the underlying simulator or training directly on sufficiently rich experimental data may be as important as further reducing surrogate error. These comparisons should nevertheless be interpreted with care, since each distribution of HoF-generated responses is compared with only one or two experimental recordings per cell. In addition, derived features can amplify small waveform discrepancies. AP width, for example, is measured at a fixed voltage threshold, so a modest error in spike amplitude or shape can induce a larger error in the estimated width.

\subsection{Preserving Cell-Type-Specific Electrophysiological Feature Distributions}\label{sec:distributions}

\looseness=-1 Figure~\ref{fig:efeature_distributions} evaluates whether \texttt{NOBLE} preserves electrophysiological feature distributions, rather than only individual voltage traces. This provides a stronger test of population-level fidelity, since a useful surrogate should reproduce variability across cells, model configurations, and input amplitudes. Across the full PVALB population, \texttt{NOBLE} closely matches the PDE-derived distributions for inverse ISI, mean firing frequency, action potential width and amplitude, and afterhyperpolarization (AHP) depth. For a representative PVALB neuron, \texttt{NOBLE} also reproduces feature distributions across stimulus amplitudes, remaining well aligned with both PDE simulations and experimental recordings. This indicates that \texttt{NOBLE} preserves both spiking temporal structure and waveform-level dynamics.

\looseness=-1 The summary statistics in Figure~\ref{fig:efeature_distributions}C make this agreement especially clear. \texttt{NOBLE} achieves strikingly high explained variability across the PDE-derived feature distributions, indicating that it captures not only the mean behavior but also the cell-to-cell and stimulus-dependent structure. This is particularly remarkable given that these electrophysiological features depend on nonlinear events such as spike initiation, repolarization, and post-spike recovery. The variance ratios and normalized mean biases remain generally close to their ideal values, showing that \texttt{NOBLE} preserves both the scale and location of the feature distributions. Together, these results demonstrate that \texttt{NOBLE} is not merely a fast voltage predictor, but a faithful population-level surrogate for biologically grounded neuronal simulation.

\vspace{-0.7mm}

\subsection{Capturing Spike Timing}\label{sec:spike_timing}

\vspace{-1mm}

\begin{figure}[tb]
    \vspace{-2mm}
    \centering
    \includegraphics[width=\linewidth]{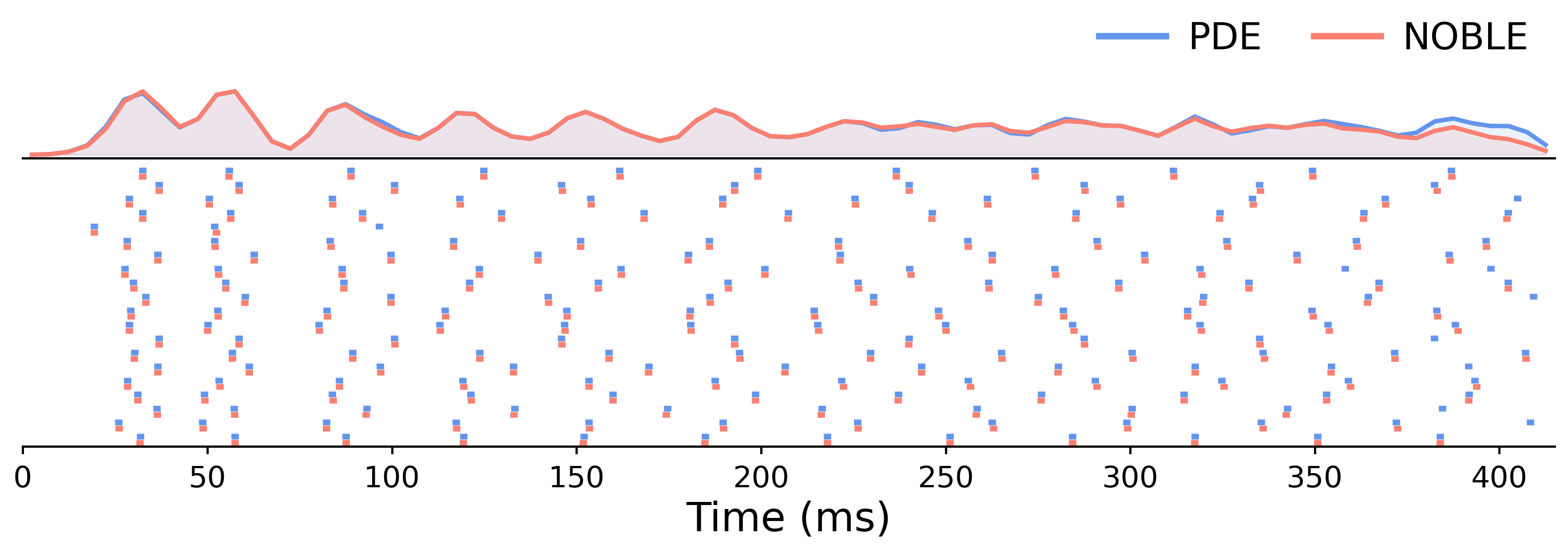}
    \vspace{-6mm}\caption{\textbf{Spike timing in PVALB neurons.} Raster comparison of spike times for a PVALB neuron.
        The top panel shows the smoothed population spike-time density for HoF simulations (blue) and \texttt{NOBLE} predictions (red).
        The bottom panel shows individual spike events across traces.
        Rasters for SST and VIP neurons are shown in \Cref{fig:rasters_sst_vip}.}
    \label{fig:rasters}
    \vspace{1mm}
\end{figure}

\looseness=-1 Accurately reproducing spike times is especially challenging because spiking is governed by sharp threshold-crossing events. Small errors in the predicted voltage trajectory, local membrane dynamics, or recovery phase can accumulate into substantial shifts in action potential onset, so a model may achieve low trajectory error while failing to preserve the temporal structure of spiking. \texttt{NOBLE} nevertheless closely matches the timing of individual spikes across all three cell types, indicating that it captures the nonlinear dynamics that determine when action potentials occur, not only the coarse membrane response. Raster plots over populations of traces show strong alignment of HoF and \texttt{NOBLE} spike events at both the single-trace and population levels (\Cref{fig:rasters,fig:rasters_sst_vip}), and representative traces preserve individual spike timing (\Cref{fig:spike-timing-examples}).

\begin{table}[t]
    \centering
    \caption{Relative $\mathrm{L2}$ test errors of the PVALB \texttt{NOBLE} model.
        \textbf{(a)} Cell-specific versus cell-type-level \texttt{NOBLE} models on one PVALB neuron.
        \textbf{(b)} Performance stratified by firing regime.
        \textbf{(c)} Out-of-distribution performance on unseen HoF parameterizations from cells represented during training (HoFs) and on cells excluded entirely from training (Cells).}
    \vspace{1mm}
    \label{tab:analysis}
    \sublabel{tab:noble_comparison}{a}
    \sublabel{tab:firing_regimes}{b}
    \sublabel{tab:out_of_distribution}{c}
    \renewcommand{\arraystretch}{1.02}
    \setlength{\tabcolsep}{5pt}
    \resizebox{\linewidth}{!}{%
        \begin{tabular}{@{}lcccccccc@{}}
            \toprule
            \textbf{Setting}
            & \textbf{Voltage}
            & \shortstack{\textbf{inv ISI}\\\textbf{values}}
            & \shortstack{\textbf{mean AP}\\\textbf{amplitude}}
            & \shortstack{\textbf{steady-state}\\\textbf{voltage}}
            & \shortstack{\textbf{AP}\\\textbf{width}}
            & \shortstack{\textbf{Spikecount}}
            & \shortstack{\textbf{Missed}\\\textbf{firing}}
            & \shortstack{\textbf{Missed}\\\textbf{non-firing}}\\
            \midrule
            \multicolumn{9}{@{}l}{\textit{(a) Model}}\\
            Cell-specific   & 1.92\% & 0.19\% & 8.46\% & 0.94\% & 7.89\% & 2.93\% & 4.40\% & 0.00\% \\
            Cell-type-level & 2.30\% & 0.55\% & 7.20\% & 0.66\% & 8.03\% & 6.03\% & 2.79\% & 0.00\% \\
            \midrule
            \multicolumn{9}{@{}l}{\textit{(b) Regime}}\\
            Threshold   & 0.55\% & 28.05\% & 7.95\% & 0.16\% & 11.77\% & 13.11\% & 9.26\% & 0.01\% \\
            Low firing  & 4.38\% & 10.92\% & 7.65\% & 0.68\% & 8.87\%  & 15.94\% & 0.00\% & 0.00\% \\
            High firing & 4.82\% & 18.49\% & 7.68\% & 0.60\% & 9.15\%  & 21.75\% & 0.02\% & 0.00\% \\
            \midrule
            \multicolumn{9}{@{}l}{\textit{(c) Out of Distribution}}\\
            HoFs  & 13.72\% & 3.76\%  & 9.42\%  & 1.36\% & 11.03\% & 11.65\% & 3.98\% & 0.00\% \\
            Cells & 7.87\%  & 75.29\% & 17.86\% & 1.75\% & 23.30\% & 52.96\% & 0.55\% & 4.01\% \\
            \bottomrule
        \end{tabular}} \vspace{3mm}
\end{table}

\subsection{Analysis of the Shared Cell-Type Operator}\label{sec:analysis}

We next probe the shared PVALB operator in three ways (\Cref{tab:analysis}), comparing it with a cell-specific surrogate, evaluating it across the firing regimes a neuron passes through as input current increases, and testing it on HoF models and cells held out from training.

\paragraph{Cell-type-level versus cell-specific training.}
\looseness=-1 By learning a shared operator across neurons within a transcriptomic cell type, \texttt{NOBLE} avoids separate cell-specific surrogates while capturing cell-level variability through electrophysiological conditioning. This population-level formulation enables interpolation between observed neurons and generation of synthetic cell-like responses within the learned cell-type manifold. A natural question is whether sharing a single operator across cells comes at the cost of per-cell predictive accuracy. We therefore compare its predictions on a PVALB neuron with those of a cell-specific \texttt{NOBLE} model trained on that neuron alone (Table~\ref{tab:noble_comparison}). The cell-specific model achieves a lower voltage-trace error ($1.92\%$ versus $2.30\%$), as well as lower inverse-ISI, AP-width, and spike-count errors. The shared model, in turn, achieves lower mean-AP-amplitude, steady-state-voltage, and missed-firing errors. Hence, sharing a single operator across multiple cells introduces only a modest reduction in per-cell accuracy while avoiding the need to train a separate surrogate for each neuron.

\paragraph{Firing regimes.}
\looseness=-1 Increasing the input current amplitude moves a neuron from a non-firing regime through threshold, low-firing, and high-firing regimes. Near threshold, small errors can determine whether a spike occurs at all, whereas at higher amplitudes the model must reproduce repeated spike timing and waveform structure. We therefore stratify PVALB test performance by regime (Table~\ref{tab:firing_regimes}), defining regimes per cell because HoF firing thresholds differ across cells. The threshold regime spans amplitudes between the minimum and maximum HoF firing thresholds of a cell, the low-firing regime runs from $0.05\,\mathrm{nA}$ above the maximum threshold up to $0.22\,\mathrm{nA}$, and the high-firing regime lies above $0.22\,\mathrm{nA}$. Waveform reconstruction is stable across regimes, with mean-AP-amplitude errors of $7.65$--$7.95\%$ and AP-width errors of $8.87$--$11.77\%$. Spike initiation is hardest at threshold, where the missed-firing rate rises to $9.26\%$, compared with at most $0.02\%$ elsewhere. Inverse-ISI error also peaks there at $28.05\%$, because traces contain few spikes and modest temporal shifts produce large relative interval errors. Voltage error, in contrast, is lowest at threshold ($0.55\%$), since trajectories remain close to the subthreshold state.

\paragraph{Out-of-distribution generalization.}
\looseness=-1 Having characterized performance across firing regimes for in-distribution test examples, we next evaluate two forms of distribution shift on PVALB neurons: held-out HoF parameterizations from cells represented during training, and entire cells excluded from training. We first consider held-out HoF parameterizations from cells that remain represented in the training set. This setting contains 5,000 test samples. As shown in Table~\ref{tab:out_of_distribution}, \texttt{NOBLE} retains low errors for several spike-timing and waveform metrics, including inverse ISI ($3.76\%$), mean AP amplitude ($9.42\%$), AP width ($11.03\%$), and spike count ($11.65\%$). The pointwise voltage error is substantially larger ($13.72\%$). This metric is highly sensitive to temporal phase error. A small displacement of one spike can shift subsequent spikes and produce a large trajectory-level $\mathrm{L2}$ error even when interspike intervals and waveform structure are recovered. Representative examples are provided in \Cref{fig:out_of_distribution_examples}.

\looseness=-1 We next consider the more challenging setting in which entire cells are excluded from training. We additionally hold out two complete PVALB cells, each with $60$ HoF models, giving $60{,}000$ test samples. On these unseen cells, the voltage error of $7.87\%$ stays below both the PVALB cell-to-cell variability of $8.34\%$ (\Cref{tab:cell_to_cell}) and the PDE-to-experiment discrepancy of $12.97\%$ (\Cref{tab:PDE_experiment}), and steady-state-voltage error and missed-firing rate remain low at $1.75\%$ and $0.55\%$. However, finer spiking statistics degrade substantially, with inverse-ISI, AP-width, and spike-count errors rising to $75.29\%$, $23.30\%$, and $52.96\%$. \texttt{NOBLE} thus retains coarse voltage and excitability characteristics for unseen cells, but generalization of detailed spiking dynamics remains limited when the target cell is not represented during training.

\begin{table}[t] \vspace{-2mm}
    \centering
    \caption{Embedding ablations for PVALB \texttt{NOBLE} (relative $\mathrm{L2}$ test errors).
        \textbf{(a)} Number of biologically informed features, added in the order \texttt{intercept}, \texttt{Spikecount}, \texttt{inv\_first\_ISI}, \texttt{AHP\_depth}, \texttt{mean\_AP\_amplitude}, \texttt{voltage\_base}, and \texttt{steady\_state\_voltage\_stimend}.
        The final model uses the first six.
        \textbf{(b)} Original two-dimensional embedding $(I_{\mathrm{thr}},s_{\mathrm{thr}})$ versus the six-feature population-level embedding.
        \textbf{(c)} Fixed versus learned feature transformations.}
    \vspace{1mm}
    \label{tab:ablations}
    \sublabel{tab:embedding_ablation}{a}
    \sublabel{tab:embedding_comparison}{b}
    \sublabel{tab:learned_embedding}{c}
    \renewcommand{\arraystretch}{1.02}
    \setlength{\tabcolsep}{5pt}
    \resizebox{\linewidth}{!}{%
        \begin{tabular}{@{}lcccccccc@{}}
            \toprule
            \textbf{Variant}
            & \textbf{Voltage}
            & \shortstack{\textbf{inv ISI}\\\textbf{values}}
            & \shortstack{\textbf{mean AP}\\\textbf{amplitude}}
            & \shortstack{\textbf{steady-state}\\\textbf{voltage}}
            & \shortstack{\textbf{AP}\\\textbf{width}}
            & \shortstack{\textbf{Spikecount}}
            & \shortstack{\textbf{Missed}\\\textbf{firing}}
            & \shortstack{\textbf{Missed}\\\textbf{non-firing}}\\
            \midrule
            \multicolumn{9}{@{}l}{\textit{(a) Features}}\\
            None & 8.01\% & -       & -       & 3.85\% & -       & -       & 100\%  & 0.00\% \\
            2    & 2.21\% & 46.21\% & 12.20\% & 0.37\% & 12.30\% & 21.66\% & 5.81\% & 0.02\% \\
            4    & 2.03\% & 44.24\% & 12.39\% & 0.24\% & 12.33\% & 22.57\% & 2.15\% & 0.00\% \\
            6    & 1.73\% & 16.97\% & 8.73\% & 0.25\% & 10.54\% & 19.27\% & 1.87\% & 0.00\% \\
            7    & 1.79\% & 17.48\% & 10.86\% & 0.25\% & 12.11\% & 20.54\% & 2.06\% & 0.00\% \\
            \midrule
            \multicolumn{9}{@{}l}{\textit{(b) Features}}\\
            2 & 2.00\% & 50.53\% & 11.49\% & 0.35\% & 11.87\% & 19.03\% & 6.47\% & 0.01\% \\
            6 & 1.73\% & 16.97\% & 8.73\%  & 0.25\% & 10.54\% & 19.27\% & 1.87\% & 0.00\% \\
            \midrule
            \multicolumn{9}{@{}l}{\textit{(c) Embedding}}\\
            Fixed   & 1.73\% & 16.97\% & 8.73\% & 0.25\% & 10.54\% & 19.27\% & 1.87\% & 0.00\% \\
            Learned & 1.69\% & 16.77\% & 9.78\% & 0.24\% & 11.36\% & 18.97\% & 1.80\% & 0.00\% \\
            \bottomrule
        \end{tabular}} 
\end{table}

\subsection{Biologically Informed Latent Embeddings}\label{sec:embeddings_ablation}

\vspace{-1mm}

\looseness=-1 Although \Cref{sec:feature_selection} ranks candidate features by the within-cell-type variation they capture, it does not fix the embedding dimension. We therefore ablate the number of top-ranked features (Table~\ref{tab:embedding_ablation}). Without feature embeddings, the model fails to predict spiking responses and voltage error is $8.01\%$. Embedding two features recovers spiking behavior and reduces voltage error to $2.21\%$, but spike-timing errors remain high, with an inverse-ISI error of $46.21\%$. Across the evaluated metrics, six features yield the lowest errors, while adding a seventh feature does not consistently improve the electrophysiological metrics. The final model therefore leverages six features.

\looseness=-1 Table~\ref{tab:embedding_comparison} compares this six-feature embedding with the original two-dimensional $(I_{\mathrm{thr}}, s_{\mathrm{thr}})$ embedding of~\citet{ghafourpour2025noble}, which becomes ambiguous across neurons (\Cref{sec:feature_selection}). When trained using the original embedding, the model primarily captures information related to firing rate, such as spike count. Performance degrades for features that depend on aspects of the dynamics not captured by $(I_{\mathrm{thr}}, s_{\mathrm{thr}})$. For example, inverse ISI, which reflects the timing between spikes, worsens nearly threefold, from $16.97\%$ to $50.53\%$. Mean AP amplitude, AP width, steady-state voltage, and voltage accuracy also deteriorate, and missed firing rises from $1.87\%$ to $6.47\%$.

\looseness=-1 Finally, we examine the effects of replacing the fixed feature-based latent space with a learned latent space. In the fixed representation, each feature is independently min--max scaled before the sinusoidal encoding of \Cref{eq:embedding}. In the learned variant, we instead standardize the feature vector and apply a learned invertible linear transformation followed by a scaled sigmoid (\Cref{app:learned_embedding}). Table~\ref{tab:learned_embedding} shows that learning this transformation provides little change in predictive performance. Voltage error decreases slightly from $1.73\%$ to $1.69\%$, inverse-ISI error from $16.97\%$ to $16.77\%$, and spike-count error from $19.27\%$ to $18.97\%$. Conversely, mean-AP-amplitude error increases from $8.73\%$ to $9.78\%$ and AP-width error from $10.54\%$ to $11.36\%$. Overall, the learned transformation does not provide a consistent improvement over the fixed, biologically informed representation, which also preserves a direct correspondence between latent coordinates and electrophysiological features. However, this learned mapping is very simple, and more expressive mappings may perform better.

\subsection{Training Objective}\label{sec:loss_results}

\looseness=-1 Having examined the conditioning representation, we next isolate the effect of the second major modification to \texttt{NOBLE}: the training objective. The overfitting studies (\Cref{app:overfitting}) identified the reference-standardized absolute $\mathrm{L4}$ loss and the sigmoid-gated relative $\mathrm{L2}$ loss ($w=10$) of \Cref{eq:soft_threshold_lp_main} as promising candidates, and we trained both on the full PVALB dataset. The sigmoid-gated loss, which penalizes signal disagreement heavily in the spike regions, improves the representation of spiking features, most notably mean AP amplitude ($15.71\%$ to $8.73\%$) and AP width ($13.11\%$ to $10.54\%$), without degrading the non-spiking components of the signal. It also reduces inverse-ISI error from $29.36\%$ to $16.97\%$ and voltage error from $2.63\%$ to $1.73\%$ (full results in \Cref{tab:full_loss_function_comparison}). We therefore use the sigmoid-gated objective for all models.

\hfill 

\section{Conclusion}\label{sec:conclusion}

\looseness=-1 We introduced a cell-type-specific \texttt{NOBLE} framework that learns a single neural operator shared across the modeled neurons of a transcriptomic cell type. This is made possible by a richer, biologically interpretable electrophysiological embedding that distinguishes models with different dynamics and a spike-weighted training objective that better preserves spike timing and waveform structure. Across major human cortical inhibitory cell types, the model reproduces subthreshold dynamics, non-spiking responses, action-potential morphology, input-output firing relationships, and spike timing, despite the sensitivity of threshold-crossing dynamics to small voltage errors. Its predicted feature distributions align closely with PDE-based HoF simulations and human experimental recordings, preserving both average behavior and structured cell-to-cell and stimulus-dependent variability. Consistent with this population-level agreement, predicted spike-triggered averages and F-I curves also retain cell-type-specific signatures across inhibitory neuron classes, including the narrow spike widths and high-frequency firing profiles characteristic of PVALB interneurons. Together with a speedup of up to three orders of magnitude over the numerical solver, these results support \texttt{NOBLE} as a practical surrogate for large-scale, biologically grounded ensemble simulation. More broadly, this work shows that neural-operator surrogates can capture structured variability across related biological systems, offering a path toward scalable tools for studying cellular diversity, generating synthetic neuronal populations, and building larger-scale models of cell-type-specific computation.

\looseness=-1 \textbf{Limitations.} This work focuses on within-cell-type population modeling using the available HoF models, electrophysiological features, and current-injection protocols. Thus, \texttt{NOBLE} should be viewed not as a replacement for detailed biophysical simulators, but as a scalable surrogate that inherits biological grounding from those simulators and the experimental data used to construct them. Future work should extend the framework to arbitrary time-varying stimuli and richer protocols, evaluate generalization across additional transcriptomic cell types, brain regions, and species, and explore learned latent embeddings that adapt the conditioning geometry while preserving biological interpretability.

\newpage

\begin{ack}
A.A. is supported by the Bren Endowed Chair, ONR (MURI grant N00014-23-1-2654), and the AI2050 Senior Fellow program at Schmidt Sciences. C.A.A. is supported by the National Institutes of Health R01-NS120300 and R01-NS130126. P.H.W. is supported by the National Institutes of Health R01-NS130126. C.Z. is supported by the Caltech Samuel P. and Frances Krown SURF Fellowship.

\end{ack}

\bibliographystyle{unsrtnat}
\bibliography{References.bib}

@article{pronneke2015characterizing,
  title={Characterizing VIP neurons in the barrel cortex of VIPcre/tdTomato mice reveals layer-specific differences},
  author={Pr{\"o}nneke, Alvar and Scheuer, Bianca and Wagener, Robin J and M{\"o}ck, Martin and Witte, Mirko and Staiger, Jochen F},
  journal={Cerebral cortex},
  volume={25},
  number={12},
  pages={4854--4868},
  year={2015},
  publisher={Oxford University Press}
}

@article{veit2023cortical,
  title={Cortical VIP neurons locally control the gain but globally control the coherence of gamma band rhythms},
  author={Veit, Julia and Handy, Gregory and Mossing, Daniel P and Doiron, Brent and Adesnik, Hillel},
  journal={Neuron},
  volume={111},
  number={3},
  pages={405--417},
  year={2023},
  publisher={Elsevier}
}

@article{pi2013cortical,
  title={Cortical interneurons that specialize in disinhibitory control},
  author={Pi, Hyun-Jae and Hangya, Bal{\'a}zs and Kvitsiani, Duda and Sanders, Joshua I and Huang, Z Josh and Kepecs, Adam},
  journal={Nature},
  volume={503},
  number={7477},
  pages={521--524},
  year={2013},
  publisher={Nature Publishing Group UK London}
}

@book{koch2004biophysics,
  title={Biophysics of computation: information processing in single neurons},
  author={Koch, Christof},
  year={2004},
  publisher={Oxford university press}
}

@article{kim2015cortically,
  title={Cortically projecting basal forebrain parvalbumin neurons regulate cortical gamma band oscillations},
  author={Kim, Tae and Thankachan, Stephen and McKenna, James T and McNally, James M and Yang, Chun and Choi, Jee Hyun and Chen, Lichao and Kocsis, Bernat and Deisseroth, Karl and Strecker, Robert E and others},
  journal={Proceedings of the National Academy of Sciences},
  volume={112},
  number={11},
  pages={3535--3540},
  year={2015},
  publisher={National Academy of Sciences}
}

@article{douglas1995recurrent,
  title={Recurrent excitation in neocortical circuits},
  author={Douglas, Rodney J and Koch, Christof and Mahowald, Misha and Martin, Kevan AC and Suarez, Humbert H},
  journal={Science},
  volume={269},
  number={5226},
  pages={981--985},
  year={1995},
  publisher={American Association for the Advancement of Science}
}

@article{buzsaki2010neural,
  title={Neural syntax: cell assemblies, synapsembles, and readers},
  author={Buzs{\'a}ki, Gy{\"o}rgy},
  journal={Neuron},
  volume={68},
  number={3},
  pages={362--385},
  year={2010},
  publisher={Elsevier}
}

@book{miettinen1999nonlinear,
  title={Nonlinear multiobjective optimization},
  author={Miettinen, Kaisa},
  volume={12},
  year={1999},
  publisher={Springer Science \& Business Media}
}

@article{achard2006complex,
  title={Complex parameter landscape for a complex neuron model},
  author={Achard, Pablo and De Schutter, Erik},
  journal={PLoS computational biology},
  volume={2},
  number={7},
  pages={e94},
  year={2006},
  publisher={Public Library of Science San Francisco, USA}
}

@article{luo2021architectures,
  title={Architectures of neuronal circuits},
  author={Luo, Liqun},
  journal={Science},
  volume={373},
  number={6559},
  pages={eabg7285},
  year={2021},
  publisher={American Association for the Advancement of Science}
}

@article{siletti2023transcriptomic,
  title={Transcriptomic diversity of cell types across the adult human brain},
  author={Siletti, Kimberly and Hodge, Rebecca and Mossi Albiach, Alejandro and Lee, Ka Wai and Ding, Song-Lin and Hu, Lijuan and L{\"o}nnerberg, Peter and Bakken, Trygve and Casper, Tamara and Clark, Michael and others},
  journal={Science},
  volume={382},
  number={6667},
  pages={eadd7046},
  year={2023},
  publisher={American Association for the Advancement of Science}
}

@article{zhang2025cell,
  title={The cell-type underpinnings of the human functional cortical connectome},
  author={Zhang, Xi-Han and Anderson, Kevin M and Dong, Hao-Ming and Chopra, Sidhant and Dhamala, Elvisha and Emani, Prashant S and Gerstein, Mark B and Margulies, Daniel S and Holmes, Avram J},
  journal={Nature Neuroscience},
  volume={28},
  number={1},
  pages={150--160},
  year={2025},
  publisher={Nature Publishing Group US New York}
}

@article{chartrand2023morphoelectric,
  title={Morphoelectric and transcriptomic divergence of the layer 1 interneuron repertoire in human versus mouse neocortex},
  author={Chartrand, Thomas and Dalley, Rachel and Close, Jennie and Goriounova, Natalia A and Lee, Brian R and Mann, Rusty and Miller, Jeremy A and Molnar, Gabor and Mukora, Alice and Alfiler, Lauren and others},
  journal={Science},
  volume={382},
  number={6667},
  pages={eadf0805},
  year={2023},
  publisher={American Association for the Advancement of Science}
}

@article{sunkin2012allen,
  title={Allen Brain Atlas: an integrated spatio-temporal portal for exploring the central nervous system},
  author={Sunkin, Susan M and Ng, Lydia and Lau, Chris and Dolbeare, Tim and Gilbert, Terri L and Thompson, Carol L and Hawrylycz, Michael and Dang, Chinh},
  journal={Nucleic acids research},
  volume={41},
  number={D1},
  pages={D996--D1008},
  year={2012},
  publisher={Oxford University Press}
}

@article{berg2021human,
  title={Human neocortical expansion involves glutamatergic neuron diversification},
  author={Berg, Jim and Sorensen, Staci A and Ting, Jonathan T and Miller, Jeremy A and Chartrand, Thomas and Buchin, Anatoly and Bakken, Trygve E and Budzillo, Agata and Dee, Nick and Ding, Song-Lin and others},
  journal={Nature},
  volume={598},
  number={7879},
  pages={151--158},
  year={2021},
  publisher={Nature Publishing Group UK London}
}

@article{lee2023signature,
  title={Signature morphoelectric properties of diverse GABAergic interneurons in the human neocortex},
  author={Lee, Brian R and Dalley, Rachel and Miller, Jeremy A and Chartrand, Thomas and Close, Jennie and Mann, Rusty and Mukora, Alice and Ng, Lindsay and Alfiler, Lauren and Baker, Katherine and others},
  journal={Science},
  volume={382},
  number={6667},
  pages={eadf6484},
  year={2023},
  publisher={American Association for the Advancement of Science}
}

@article{gouwens2020integrated,
  title={Integrated morphoelectric and transcriptomic classification of cortical GABAergic cells},
  author={Gouwens, Nathan W and Sorensen, Staci A and Baftizadeh, Fahimeh and Budzillo, Agata and Lee, Brian R and Jarsky, Tim and Alfiler, Lauren and Baker, Katherine and Barkan, Eliza and Berry, Kyla and others},
  journal={Cell},
  volume={183},
  number={4},
  pages={935--953},
  year={2020},
  publisher={Elsevier}
}

@article{druckmann2007novel,
  title={A novel multiple objective optimization framework for constraining conductance-based neuron models by experimental data},
  author={Druckmann, Shaul and Banitt, Yoav and Gidon, Albert A and Sch{\"u}rmann, Felix and Markram, Henry and Segev, Idan},
  journal={Frontiers in neuroscience},
  volume={1},
  pages={56},
  year={2007},
  publisher={Frontiers}
}

@article{reimann2013biophysically,
  title={A biophysically detailed model of neocortical local field potentials predicts the critical role of active membrane currents},
  author={Reimann, Michael W and Anastassiou, Costas A and Perin, Rodrigo and Hill, Sean L and Markram, Henry and Koch, Christof},
  journal={Neuron},
  volume={79},
  number={2},
  pages={375--390},
  year={2013},
  publisher={Elsevier}
}

@article{buchin2022multi,
  title={Multi-modal characterization and simulation of human epileptic circuitry},
  author={Buchin, Anatoly and de Frates, Rebecca and Nandi, Anirban and Mann, Rusty and Chong, Peter and Ng, Lindsay and Miller, Jeremy and Hodge, Rebecca and Kalmbach, Brian and Bose, Soumita and others},
  journal={Cell reports},
  volume={41},
  number={13},
  year={2022},
  publisher={Elsevier}
}

@article{mosher2020cellular,
  title={Cellular classes in the human brain revealed in vivo by heartbeat-related modulation of the extracellular action potential waveform},
  author={Mosher, Clayton P and Wei, Yina and Kami{\'n}ski, Jan and Nandi, Anirban and Mamelak, Adam N and Anastassiou, Costas A and Rutishauser, Ueli},
  journal={Cell reports},
  volume={30},
  number={10},
  pages={3536--3551},
  year={2020},
  publisher={Elsevier}
}

@article{van2016bluepyopt,
  title={BluePyOpt: leveraging open source software and cloud infrastructure to optimise model parameters in neuroscience},
  author={Van Geit, Werner and Gevaert, Michael and Chindemi, Giuseppe and R{\"o}ssert, Christian and Courcol, Jean-Denis and Muller, Eilif B and Sch{\"u}rmann, Felix and Segev, Idan and Markram, Henry},
  journal={Frontiers in neuroinformatics},
  volume={10},
  pages={17},
  year={2016},
  publisher={Frontiers Media SA}
}

@article{gratiy2018bionet,
  title={BioNet: A Python interface to NEURON for modeling large-scale networks},
  author={Gratiy, Sergey L and Billeh, Yazan N and Dai, Kael and Mitelut, Catalin and Feng, David and Gouwens, Nathan W and Cain, Nicholas and Koch, Christof and Anastassiou, Costas A and Arkhipov, Anton},
  journal={PloS one},
  volume={13},
  number={8},
  pages={e0201630},
  year={2018},
  publisher={Public Library of Science San Francisco, CA USA}
}

@article{hines2001neuron,
  title={NEURON: a tool for neuroscientists},
  author={Hines, Michael L and Carnevale, Nicholas T},
  journal={The neuroscientist},
  volume={7},
  number={2},
  pages={123--135},
  year={2001},
  publisher={Sage Publications Sage CA: Thousand Oaks, CA}
}

@article{nandi2022single,
  title={Single-neuron models linking electrophysiology, morphology, and transcriptomics across cortical cell types},
  author={Nandi, Anirban and Chartrand, Thomas and Van Geit, Werner and Buchin, Anatoly and Yao, Zizhen and Lee, Soo Yeun and Wei, Yina and Kalmbach, Brian and Lee, Brian and Lein, Ed and others},
  journal={Cell reports},
  volume={40},
  number={6},
  year={2022},
  publisher={Elsevier}
}

@article{su2021deep,
  title={Deep learning of biological models from data: applications to ODE models},
  author={Su, Wei-Hung and Chou, Ching-Shan and Xiu, Dongbin},
  journal={Bulletin of mathematical biology},
  volume={83},
  pages={1--19},
  year={2021},
  publisher={Springer}
}

@article{kuptsov2023discovering,
  title={Discovering dynamical features of Hodgkin--Huxley-type model of physiological neuron using artificial neural network},
  author={Kuptsov, Pavel V and Stankevich, Nataliya V and Bagautdinova, Elmira R},
  journal={Chaos, Solitons \& Fractals},
  volume={167},
  pages={113027},
  year={2023},
  publisher={Elsevier}
}

@article{ferrante2022physically,
  title={Physically constrained neural networks to solve the inverse problem for neuron models},
  author={Ferrante, Matteo and Duggento, Andera and Toschi, Nicola},
  journal={arXiv preprint arXiv:2209.11998},
  year={2022}
}

@article{fitzhugh1955mathematical,
  title={Mathematical models of threshold phenomena in the nerve membrane},
  author={FitzHugh, Richard},
  journal={The bulletin of mathematical biophysics},
  volume={17},
  pages={257--278},
  year={1955},
  publisher={Springer}
}

@article{nagumo1962active,
  title={An active pulse transmission line simulating nerve axon},
  author={Nagumo, Jinichi and Arimoto, Suguru and Yoshizawa, Shuji},
  journal={Proceedings of the IRE},
  volume={50},
  number={10},
  pages={2061--2070},
  year={1962},
  publisher={IEEE}
}

@article{hodgkin1952quantitative,
  title={A quantitative description of membrane current and its application to conduction and excitation in nerve},
  author={Hodgkin, Alan L and Huxley, Andrew F},
  journal={The Journal of physiology},
  volume={117},
  number={4},
  pages={500},
  year={1952}
}

@article{pandey2024efficient,
  title={An efficient wavelet-based physics-informed neural networks for singularly perturbed problems},
  author={Pandey, Himanshu and Singh, Anshima and Behera, Ratikanta},
  journal={arXiv preprint arXiv:2409.11847},
  year={2024}
}

@inproceedings{werneck2023replacing,
  title={Replacing the FitzHugh-Nagumo Electrophysiology Model by Physics-Informed Neural Networks},
  author={Werneck, Yan Barbosa and dos Santos, Rodrigo Weber and Rocha, Bernardo Martins and Oliveira, Rafael Sachetto},
  booktitle={International Conference on Computational Science},
  pages={699--713},
  year={2023},
  organization={Springer}
}

@inproceedings{rudi2022parameter,
  title={Parameter estimation with dense and convolutional neural networks applied to the FitzHugh--Nagumo ODE},
  author={Rudi, Johann and Bessac, Julie and Lenzi, Amanda},
  booktitle={Mathematical and Scientific Machine Learning},
  pages={781--808},
  year={2022},
  organization={PMLR}
}

@article{centofanti2024learning,
  title={Learning the Hodgkin--Huxley model with operator learning techniques},
  author={Centofanti, Edoardo and Ghiotto, Massimiliano and Pavarino, Luca F},
  journal={Computer Methods in Applied Mechanics and Engineering},
  volume={432},
  pages={117381},
  year={2024},
  publisher={Elsevier}
}

@article{Raissi2019,
title = {Physics-informed neural networks: A deep learning framework for solving forward and inverse problems involving nonlinear partial differential equations},
journal = {Journal of Computational Physics},
volume = {378},
pages = {686-707},
year = {2019},
issn = {0021-9991},
doi = {10.1016/j.jcp.2018.10.045},
author = {M. Raissi and P. Perdikaris and G.E. Karniadakis},
}

@article{Raissi2017Part1,
  title={Physics Informed Deep Learning (Part I): Data-driven Solutions of Nonlinear Partial Differential Equations},
  author={Maziar Raissi and Paris Perdikaris and George E. Karniadakis},
  journal={ArXiv},
  year={2017},
  volume={abs/1711.10561}
}

@article{Raissi2017Part2,
  title={Physics Informed Deep Learning (Part II): Data-driven Discovery of Nonlinear Partial Differential Equations},
  author={Maziar Raissi and Paris Perdikaris and George E. Karniadakis},
  journal={ArXiv},
  year={2017},
  volume={abs/1711.10566}
}

@article{Kovachki2021_2,
author = {Kovachki, Nikola and Lanthaler, Samuel and Mishra, Siddhartha},
title = {On Universal Approximation and Error Bounds for {F}ourier Neural Operators},
year = {2021},
publisher = {JMLR.org},
volume = {22},
number = {1},
issn = {1532-4435},
journal = {J. Mach. Learn. Res.},
articleno = {290},
}

@article{kovachki2023neural,
  title={Neural operator: Learning maps between function spaces with applications to pdes},
  author={Kovachki, Nikola and Li, Zongyi and Liu, Burigede and Azizzadenesheli, Kamyar and Bhattacharya, Kaushik and Stuart, Andrew and Anandkumar, Anima},
  journal={Journal of Machine Learning Research},
  volume={24},
  number={89},
  pages={1--97},
  year={2023}
}

@article{azizzadenesheli2024neural,
  title={Neural operators for accelerating scientific simulations and design},
  author={Azizzadenesheli, Kamyar and Kovachki, Nikola and Li, Zongyi and Liu-Schiaffini, Miguel and Kossaifi, Jean and Anandkumar, Anima},
  journal={Nature Reviews Physics},
  pages={1--9},
  year={2024},
  publisher={Nature Publishing Group UK London}
}

@article{li2020fno,
  title={Fourier neural operator for parametric partial differential equations},
  author={Li, Zongyi and Kovachki, Nikola and Azizzadenesheli, Kamyar and Liu, Burigede and Bhattacharya, Kaushik and Stuart, Andrew and Anandkumar, Anima},
  journal={arXiv preprint arXiv:2010.08895},
  year={2020}
}

@misc{lin2025mGNO,
      title={Enabling Automatic Differentiation with Mollified Graph Neural Operators}, 
      author={Ryan Y. Lin and Julius Berner and Valentin Duruisseaux and David Pitt and Daniel Leibovici and Jean Kossaifi and Kamyar Azizzadenesheli and Anima Anandkumar},
      year={2025},
      eprint={2504.08277},
      archivePrefix={arXiv},
      primaryClass={cs.LG},
}

@article{li2024pino,
  title={Physics-informed neural operator for learning partial differential equations},
  author={Li, Zongyi and Zheng, Hongkai and Kovachki, Nikola and Jin, David and Chen, Haoxuan and Liu, Burigede and Azizzadenesheli, Kamyar and Anandkumar, Anima},
  journal={ACM/JMS Journal of Data Science},
  volume={1},
  number={3},
  pages={1--27},
  year={2024},
  publisher={ACM New York, NY}
}

@Article{Wen2023,
author = {Wen, Gege and Li, Zongyi and Long, Qirui and Azizzadenesheli, Kamyar and Anandkumar, Anima and Benson, Sally M.},
title  = {Real-time high-resolution {CO2} geological storage prediction using nested {F}ourier neural operators},
journal  ={Energy Environ. Sci.},
year  ={2023},
volume  ={16},
issue  ={4},
pages  ={1732-1741},
publisher  ={The Royal Society of Chemistry},
doi  = {10.1039/D2EE04204E},
}

@article{Kurth2022,
author = {Kurth, Thorsten and Subramanian, Shashank and Harrington, Peter and Pathak, Jaideep and Mardani, Morteza and Hall, David and Miele, Andrea and Kashinath, Karthik and Anandkumar, Animashree},
year = {2022},
title = {{FourCastNet}: Accelerating Global High-Resolution Weather Forecasting using Adaptive {F}ourier Neural Operators},
doi = {10.48550/arXiv.2208.05419}
}

@Article{Berner2025Principles,
      title={Principled Approaches for Extending Neural Architectures to Function Spaces for Operator Learning}, 
      author={Julius Berner and Miguel Liu-Schiaffini and Jean Kossaifi and Valentin Duruisseaux and Boris Bonev and Kamyar Azizzadenesheli and Anima Anandkumar},
      year={2025},
      eprint={2506.10973},
      archivePrefix={arXiv},
      primaryClass={cs.LG},
}

@misc{duruisseaux2025FNOGuide,
      title={Fourier Neural Operators Explained: A Practical Perspective}, 
      author={Valentin Duruisseaux and Jean Kossaifi and Anima Anandkumar},
      year={2025},
      eprint={2512.01421},
      archivePrefix={arXiv},
      primaryClass={cs.LG},
}

@article{gopakumar2023fourier,
  title={Fourier neural operator for plasma modelling},
  author={Gopakumar, Vignesh and Pamela, Stanislas and Zanisi, Lorenzo and Li, Zongyi and Anandkumar, Anima and Team, MAST},
  journal={arXiv preprint arXiv:2302.06542},
  year={2023}
}

@misc{ganeshram2025fcpinohighprecisionphysicsinformed,
      title={FC-PINO: High Precision Physics-Informed Neural Operators via Fourier Continuation}, 
      author={Adarsh Ganeshram and Haydn Maust and Valentin Duruisseaux and Zongyi Li and Yixuan Wang and Daniel Leibovici and Oscar Bruno and Thomas Hou and Anima Anandkumar},
      year={2025},
      eprint={2211.15960},
      archivePrefix={arXiv},
      primaryClass={cs.LG},
}

@misc{kossaifi2024neural,
   title={A Library for Learning Neural Operators},
   author={Jean Kossaifi and Nikola Kovachki and Zongyi Li and David Pitt and Miguel
Liu-Schiaffini and Robert Joseph George and Boris Bonev and Kamyar
Azizzadenesheli and Julius Berner and Valentin Duruisseaux and Anima
Anandkumar},
   year={2024},
   eprint={2412.10354},
   archivePrefix={arXiv},
   primaryClass={cs.LG}
}

@inproceedings{ghafourpour2025noble,
  title={NOBLE: Neural Operator with Biologically-informed Latent Embeddings to Capture Experimental Variability in Biological Neuron Models},
  author={Luca Ghafourpour and Valentin Duruisseaux and Bahareh Tolooshams and Philip H. Wong and Costas A. Anastassiou and Anima Anandkumar},
  booktitle={Proceedings of the 39th Conference on Neural Information Processing Systems (NeurIPS 2025)},
  year={2025},
  organization={NeurIPS},
}

@article{jackson2024amygdala,
  title={Amygdala-hippocampus somatostatin interneuron beta-synchrony underlies a cross-species biomarker of emotional state},
  author={Jackson, Adam D and Cohen, Joshua L and Phensy, Aarron J and Chang, Edward F and Dawes, Heather E and Sohal, Vikaas S},
  journal={Neuron},
  volume={112},
  number={7},
  pages={1182--1195},
  year={2024},
  publisher={Elsevier}
}

@article{park2020optogenetic,
  title={Optogenetic activation of parvalbumin and somatostatin interneurons selectively restores theta-nested gamma oscillations and oscillation-induced spike timing-dependent long-term potentiation impaired by amyloid $\beta$ oligomers},
  author={Park, Kyerl and Lee, Jaedong and Jang, Hyun Jae and Richards, Blake A and Kohl, Michael M and Kwag, Jeehyun},
  journal={BMC biology},
  volume={18},
  number={1},
  pages={7},
  year={2020},
  publisher={Springer}
}

@article{deistler2025jaxley,
  title={Jaxley: differentiable simulation enables large-scale training of detailed biophysical models of neural dynamics},
  author={Deistler, Michael and Kadhim, Kyra L and Pals, Matthijs and Beck, Jonas and Huang, Ziwei and Gloeckler, Manuel and Lappalainen, Janne K and Schr{\"o}der, Cornelius and Berens, Philipp and Gon{\c{c}}alves, Pedro J and others},
  journal={Nature Methods},
  pages={1--9},
  year={2025},
  publisher={Nature Publishing Group US New York}
}

@article{mainen1995reliability,
  title={Reliability of spike timing in neocortical neurons},
  author={Mainen, Zachary F and Sejnowski, Terrence J},
  journal={Science},
  volume={268},
  number={5216},
  pages={1503--1506},
  year={1995},
  publisher={American Association for the Advancement of Science}
}

@article{mainen1996influence,
  title={Influence of dendritic structure on firing pattern in model neocortical neurons},
  author={Mainen, Zachary F and Sejnowski, Terrence J},
  journal={Nature},
  volume={382},
  number={6589},
  pages={363--366},
  year={1996},
  publisher={Nature Publishing Group UK London}
}

@article{anastassiou2011ephaptic,
  title={Ephaptic coupling of cortical neurons},
  author={Anastassiou, Costas A and Perin, Rodrigo and Markram, Henry and Koch, Christof},
  journal={Nature neuroscience},
  volume={14},
  number={2},
  pages={217--223},
  year={2011},
  publisher={Nature Publishing Group}
}

@article{lee2024cell,
  title={Cell-class-specific electric field entrainment of neural activity},
  author={Lee, Soo Yeun and Kozalakis, Konstantinos and Baftizadeh, Fahimeh and Campagnola, Luke and Jarsky, Tim and Koch, Christof and Anastassiou, Costas A},
  journal={Neuron},
  volume={112},
  number={15},
  pages={2614--2630},
  year={2024},
  publisher={Elsevier}
}

@article{tremblay2016gabaergic,
  title={GABAergic interneurons in the neocortex: from cellular properties to circuits},
  author={Tremblay, Robin and Lee, Soohyun and Rudy, Bernardo},
  journal={Neuron},
  volume={91},
  number={2},
  pages={260--292},
  year={2016},
  publisher={Elsevier}
}

\newpage 
\appendix 
\counterwithin{table}{section}
\counterwithin{figure}{section}
\section{Dataset and Hall-of-Fame Model Construction}\label[appsec]{app:hof_models}

To demonstrate the scalability of the cell-type-specific \texttt{NOBLE} framework, we utilized a Patch-seq dataset of human cortical inhibitory neurons from the middle temporal gyrus \cite{lee2023signature}. This broader Patch-seq resource contains four major transcriptomically defined inhibitory classes: parvalbumin (PVALB), somatostatin (SST), vasoactive intestinal peptide (VIP), and Lysosomal Associated Membrane Protein Family Member 5 (LAMP5). We focus on the three classes for which biophysical models were available in our dataset, whose circuit roles are summarized in \Cref{sec:background}. LAMP5 neurons are also an important inhibitory class, prominently located in the superficial layers of the human cortex, but were not included in the present modeling dataset and are therefore left for future work.

To approximate trial-to-trial variability, we generate a population of up to 60 HoF models for every individual neuron across the three transcriptomic classes. The models are instantiated via the BioNet interface \cite{gratiy2018bionet}, leveraging the NEURON simulation engine \cite{hines2001neuron} to resolve the underlying cable equations of membrane potentials into spatially discretized systems of ordinary differential equations. We employ an "all-active" configuration \cite{buchin2022multi,nandi2022single} that distributes active ion channels across somatic and dendritic compartments along the entire neuron morphology. For each neuron, we utilize a multi-objective evolutionary optimization framework \cite{nandi2022single} to find ion conductance parameter sets that faithfully replicate electrophysiological features of interest (\Cref{app:e_features_appx}). This optimization framework follows a two-stage protocol that fits passive model parameters associated with subthreshold membrane properties in the first stage, and then fits active ion channel parameters that define a neuron's firing characteristics in the second stage. Parameter sets that generate the lowest z-score error between simulated and experimental electrophysiological features are selected as HoF models over 250 generations of evolutionary optimization.

\clearpage

\section{Electrophysiological Features}\label[appsec]{app:e_features_appx}

For electrophysiological feature extraction and metrics, we use code from the Electrophys Feature Extraction Library (eFEL) available at
\begin{center}
    \url{https://github.com/BlueBrain/eFEL}
\end{center}

\vspace{5mm}

The formulas, codes, and more details about each electrophysiological feature can be found at
\begin{center}
    \url{https://efel.readthedocs.io/en/latest/eFeatures.html}
\end{center}

\vspace{5mm}

We list below the important electrophysiological features and metrics of interest when constructing neuron models (where AP denotes action potential and AHP denotes after-hyperpolarization):

{\setlength{\leftmargini}{2em}
    \setlength{\itemsep}{1em}
    \begin{itemize}
        \item \texttt{AHP\_depth}: Relative voltage values at the first AHP

        \item \texttt{AHP\_time\_from\_peak}: Time between AP peaks and first AHP depth

        \item \texttt{AHP1\_depth\_from\_peak}: Voltage difference between the first AP peak and first AHP depth

        \item \texttt{AP1\_peak}: The peak voltage of the first AP

        \item \texttt{AP1\_width}: Width of first spike at half spike amplitude, with the spike amplitude taken as the difference between the minimum between two peaks and the next peak

        \item \texttt{decay\_time\_constant\_after\_stim}: The decay time constant of the voltage right after the stimulus

        \item \texttt{depol\_block}: Checks for a depolarization block, returning 1 if there is a depolarization or hyperpolarization block and 0 otherwise

        \item \texttt{inv\_first\_ISI}: 1.0 over first interspike interval; returns 0 when no interspike interval

        \item \texttt{inv\_ISI\_values}: 1.0 over all interspike intervals

        \item \texttt{mean\_AP\_amplitude}: The mean of all of the AP amplitudes

        \item \texttt{mean\_frequency}: The mean frequency of the firing rate

        \item \texttt{peak\_time}: Time points at which peak maxima occur

        \item \texttt{sag\_amplitude}: The difference between the minimal voltage and the steady state at the end of the stimulus

        \item \texttt{spikecount}: Number of spikes in the trace, including outside of stimulus interval

        \item \texttt{steady\_state\_voltage}: The average voltage after the stimulus

        \item \texttt{steady\_state\_voltage\_stimend}: The average voltage during the last 10\% of the stimulus duration

        \item \texttt{time\_to\_first\_spike}: Time from the start of the stimulus to the maximum of the first peak

        \item \texttt{voltage\_base}: The average voltage during the last 10\% of time before the stimulus

    \end{itemize}}

The features \texttt{intercept} and \texttt{slope} are not computed with eFEL. They denote the threshold current $I_{\mathrm{thr}}$ and local threshold slope $s_{\mathrm{thr}}$ of the F-I curve (\Cref{sec:noble}).

\section{The Fourier Neural Operator}\label[appsec]{app:fno}

Neural operators~\citep{azizzadenesheli2024neural,kovachki2023neural,Berner2025Principles} provide a principled extension of neural networks for learning mappings between function spaces, and enjoy a universal operator approximation property~\citep{Kovachki2021_2}. They approximate highly nonlinear operators by composing linear integral operators $\mathcal{K}$ with pointwise nonlinear activation functions~$\sigma$. Specifically, we define the neural operator

\begin{equation}
    \mathcal{G}_{\theta} \ = \  \mathcal{Q} \ \circ \ \sigma (W_{L} + \mathcal{K}_{L} + b_L) \ \circ \  \cdots \ \circ \ \sigma(W_1 + \mathcal{K}_1 + b_1) \  \circ \ \mathcal{P}
\end{equation}

where $\mathcal{P}$ and $\mathcal{Q}$ are pointwise neural networks that lift a lower-dimensional function to a higher-dimensional representation and project it back. The model consists of $L$ stacked layers of the form $\sigma(W_{l} + \mathcal{K}_{l} + b_{l})$, where $W_l$ are pointwise linear operators, or matrices, $\mathcal{K}_l$ are integral kernel operators, $b_l $ are bias terms, and $\sigma$ denotes a fixed activation function. The parameter set $\theta$ includes all trainable parameters in $\mathcal{P}, \mathcal{Q}, W_l, \mathcal{K}_l$ and $ b_l$.

\vspace{4mm}

Kossaifi et al.\ \cite{kossaifi2024neural} provide a comprehensive open-source PyTorch library for learning neural operators, which forms the basis of our implementation. Prior knowledge of physical laws and governing differential equations can also be incorporated during training through additional loss terms, either supplementing or replacing reference data, as in physics-informed neural operators~\cite{li2024pino, lin2025mGNO, ganeshram2025fcpinohighprecisionphysicsinformed}.

\vspace{4mm}

\begin{figure}[!h]
    \vspace{8mm}
    \centering
    \includegraphics[width=\textwidth]{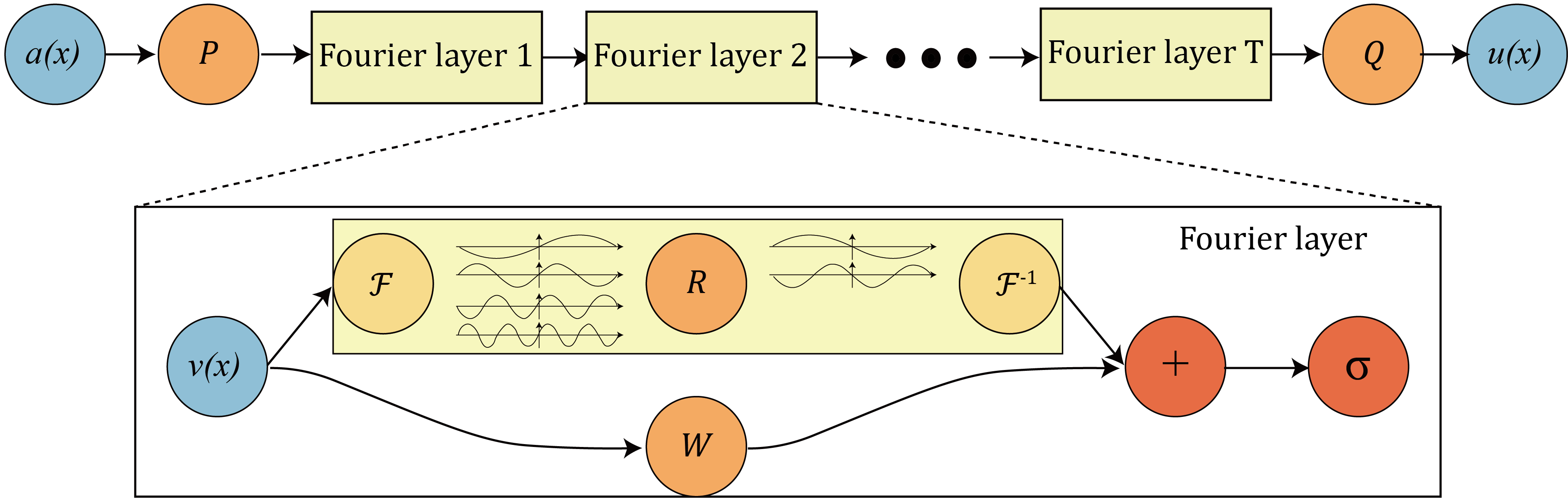}
    \caption{The Fourier Neural Operator (FNO) architecture (extracted from~\citep{li2020fno}).}
    \label{fig:fno_architecture}
\end{figure}

Several neural operator architectures have been introduced, including the Fourier Neural Operator (FNO)~\citep{li2020fno,duruisseaux2025FNOGuide}, and have been successfully applied across a broad range of problems~\citep{gopakumar2023fourier, Kurth2022, Wen2023}. An FNO (\Cref{fig:fno_architecture}) is a neural operator whose integral operator layers are defined in Fourier space as

\begin{equation}
    \label{eq:Fourier}
    \bigl(\mathcal{K}(\phi)v_t\bigr)(x)=
    \mathcal{F}^{-1}\Bigl(R_\phi \cdot (\mathcal{F} v_t) \Bigr)(x)
\end{equation}

where $R_\phi$ is the Fourier transform of a periodic function $\kappa$ parameterized by $\phi$. On a uniform mesh, the Fourier transform $\mathcal{F}$ can be computed efficiently using the fast Fourier transform (FFT). Since electrophysiological recordings and PDE simulations are sampled on equidistant time grids, this makes the FNO well suited to our setting (\Cref{sec:noble}). By operating in the frequency domain, FNOs capture global, nonlinear, and high-frequency structure in voltage dynamics, supporting generalization across temporal resolutions, input currents, and neuronal configurations without retraining.

\clearpage

\section{Embedding Analysis}\label[appsec]{app:embedding_analysis}

This appendix details the feature-selection procedure of \Cref{sec:feature_selection}.

Candidate features are extracted from simulations of each HoF model under two DC current steps of 1\,s duration, with amplitudes $0.28$\,nA and $-0.11$\,nA, chosen to probe suprathreshold spiking and subthreshold passive responses, respectively. \Cref{tab:feature_importance} reports the $\mathrm{MAD}$ scores of the main candidate features. Features with a ``.1'' suffix were extracted during the hyperpolarizing stimulus ($-0.11$\,nA), and all others during the depolarizing stimulus ($0.28$\,nA). After sorting features by decreasing $\mathrm{MAD}$ score, we sequentially retain each candidate only if its absolute correlation with all previously selected features is below a fixed threshold. The $\mathrm{MAD}$ ranking favors features that best distinguish HoF models, while correlation filtering removes dimensions that largely duplicate information already retained. We use this as a principled and interpretable selection heuristic rather than an optimality criterion: we do not claim that it identifies the feature subset that minimizes downstream prediction error. Instead, its purpose is to reduce the chance that HoF models with substantially different dynamics are assigned similar latent coordinates, without introducing an additional learned optimization step (an alternative in which the transformation of the features is learned is evaluated in \Cref{app:learned_embedding}). The number of embedded features is treated as a hyperparameter (\Cref{tab:embedding_ablation}), and the procedure yields the following embedded features for each cell type:

\begin{itemize}
    \item \textbf{PVALB:} \texttt{intercept}, \texttt{Spikecount}, \texttt{inv\_first\_ISI}, \texttt{AHP\_depth}, \texttt{mean\_AP\_amplitude}, \texttt{voltage\_base}
    \item \textbf{SST:} \texttt{AHP\_depth}, \texttt{steady\_state\_voltage}, \texttt{AP1\_peak}, \texttt{intercept}, \texttt{sag\_amplitude}, \texttt{inv\_first\_ISI}
    \item \textbf{VIP:} \texttt{voltage\_base}, \texttt{AHP\_depth}, \texttt{AP1\_peak}, \texttt{Spikecount}, \texttt{sag\_amplitude}, \texttt{inv\_first\_ISI}
\end{itemize}

\hfill 

\begin{table*}[h]
    \centering
    \caption{MAD scores for electrophysiological features across PVALB, SST, and VIP cell types.
        Scores are rounded to four decimal places.}
    \label{tab:feature_importance}
    \renewcommand{\arraystretch}{1.15}
    \setlength{\tabcolsep}{12pt}
    \begin{tabular}{@{}lccc@{}}
        \toprule
        \textbf{Feature} & \textbf{PVALB} & \textbf{SST} & \textbf{VIP} \\
        \midrule
        \texttt{AHP\_depth}                           & 0.6360 & 0.5603 & 0.5290 \\
        \texttt{AHP1\_depth\_from\_peak}              & 0.4656 & 0.5393 & 0.4653 \\
        \texttt{steady\_state\_voltage.1}             & 0.4506 & 0.5338 & 0.6006 \\
        \texttt{steady\_state\_voltage}               & 0.5005 & 0.5256 & 0.4841 \\
        \texttt{voltage\_base}                        & 0.5088 & 0.5021 & 0.5252 \\
        \texttt{voltage\_base.1}                      & 0.5169 & 0.5013 & 0.5301 \\
        \texttt{steady\_state\_voltage\_stimend}      & 0.2712 & 0.4893 & 0.3064 \\
        \texttt{AP1\_peak}                            & 0.4606 & 0.4819 & 0.5166 \\
        \texttt{decay\_time\_constant\_after\_stim.1} & 0.3609 & 0.4746 & 0.3837 \\
        \texttt{steady\_state\_voltage\_stimend.1}    & 0.4913 & 0.4535 & 0.4814 \\
        \texttt{mean\_AP\_amplitude}                  & 0.5104 & 0.4411 & 0.4971 \\
        \texttt{decay\_time\_constant\_after\_stim}   & 0.1704 & 0.4173 & 0.5042 \\
        \texttt{AP1\_width}                           & 0.2627 & 0.3599 & 0.4388 \\
        \texttt{intercept}                            & 1.0934 & 0.3197 & 0.4339 \\
        \texttt{sag\_amplitude}                       & 0.1515 & 0.3159 & 0.4319 \\
        \texttt{inv\_first\_ISI}                      & 0.6621 & 0.3019 & 0.3801 \\
        \texttt{slope}                                & 0.9184 & 0.2969 & 0.4415 \\
        \texttt{Spikecount}                           & 1.0000 & 0.2917 & 0.5000 \\
        \texttt{AHP\_time\_from\_peak}                & 0.2053 & 0.2582 & 0.2678 \\
        \texttt{time\_to\_first\_spike}               & 0.3440 & 0.1927 & 0.4083 \\
        \bottomrule
    \end{tabular}
\end{table*}

\section{Learning the Latent Embedding Space}\label[appsec]{app:learned_embedding}

The latent representation used in the main text is not learned but constructed directly from fixed electrophysiological features (\Cref{sec:feature_selection}), which serve as biologically interpretable descriptors of each HoF model. As an exploratory ablation, we test whether learning a transformation of these features improves surrogate accuracy. In the fixed representation, each feature is independently min--max scaled before the sinusoidal encoding of \Cref{eq:embedding}. For the learned variant, we instead standardize the feature vector and apply a single linear transformation followed by a sigmoid:

\begin{equation*}
    \mathbf{y} =10 \cdot \sigma\!\left(\mathbf{x}\exp(\mathbf{W})^\top + \mathbf{b} \right)+ 0.5,
\end{equation*}

where $\mathbf{x}$ denotes the standardized electrophysiological features and $\mathbf{W}$ and $\mathbf{b}$ are learned parameters. The matrix exponential ensures that the linear transformation is invertible, preventing distinct feature vectors from being collapsed by the linear map. The scaled sigmoid maps each coordinate to $[0.5,10.5]$ before applying the sinusoidal encoding.

Results are reported in Table~\ref{tab:learned_embedding} and discussed in \Cref{sec:embeddings_ablation}.

\hfill \\

\hfill 

\section{Training Objective and Evaluation Metrics}\label[appsec]{app:evaluation_metrics_appx}

\subsection{Loss Functions}

The standard objective for learning time-dependent dynamics is the relative $\mathrm{L}^p$ error:

\begin{equation}\label{eq:lp_norm}
    \text{Relative Lp error}(x, y \,; \epsilon) = \frac{\|x - y\|_p}{\|y\|_p + \epsilon} = \frac{(\sum_i |x_i - y_i|^p)^{1/p}}{(\sum_i |y_i|^p)^{1/p} + \epsilon}.
\end{equation}

\citet{ghafourpour2025noble} trained \texttt{NOBLE} with a relative $\mathrm{L4}$ loss. Because action potentials occupy only a small fraction of each trace, errors in spike timing and waveform shape contribute relatively little to this objective despite being electrophysiologically important. The spike-weighted objective of \Cref{eq:soft_threshold_lp_main} therefore adds a sigmoid-gated term that places greater weight on errors in the spiking region while retaining the original objective over the full voltage trace. In \Cref{eq:soft_threshold_lp_main}, $g(y_i)$ is a sigmoid gate that emphasizes selected voltage regimes, $w$ controls the contribution of the weighted term, and $\epsilon$ prevents division by zero.

\hfill 

\subsection{Overfitting Study}\label[appsec]{app:overfitting}

The broader diversity of dynamics in the multi-neuron setting makes the choice of training objective particularly critical. To guide it, we conducted small-scale ``overfitting'' studies on $200$ representative trajectories: \texttt{NOBLE} was trained with each candidate loss function for $2000$ epochs using the same architecture and hyperparameters as the full-scale runs, with a StepLR scheduler that reduced the learning rate every $300$ epochs. These experiments measure training error on a tiny dataset rather than test error on the full multi-neuron dataset, and therefore do not directly measure generalization. Instead, they isolate how each loss function prioritizes waveform accuracy and electrophysiological structure when data scarcity is not a confounding factor, providing a controlled diagnostic for ruling out loss functions that fail to reproduce key electrophysiological properties even when very low errors can be achieved, while identifying more promising candidates for full-scale training.

\clearpage

Table~\ref{tab:loss_function_ablation} reports the loss metrics obtained by training \texttt{NOBLE} with a collection of different loss functions, grouped as follows:

\vspace{2mm}

\begin{itemize}
    \item \textbf{Conventional and normalized $\text{Lp}$ losses.}
        This group evaluates the absolute and relative $\mathrm{Lp}$ losses defined in \Cref{eq:lp_norm}, together with variants in which the voltage traces are transformed using training-set statistics.
        We consider both standardization and min-max normalization, applied either to the reference trace alone (``ref'') or to both the prediction and reference (``both'').
        For example, for the ref-std variant, let $\mu$ and $\sigma$ denote the mean and standard deviation of the training-set reference traces and define the standardized target as
        \begin{equation}
            \tilde{y}_i = \frac{y_i-\mu}{\sigma}.
        \end{equation}

        The corresponding relative $\mathrm{Lp}$ loss is obtained by evaluating \Cref{eq:lp_norm} with $\tilde{y}$ as the target: $\mathrm{Lp}(x,\tilde{y} \,;\epsilon)$.
        The remaining variants are defined analogously using the corresponding transformation.

        \vspace{4mm}

    \item \textbf{Sigmoid-gated $\text{Lp}$ losses.}
        This group uses the sigmoid-gated objective from \Cref{eq:soft_threshold_lp_main}.

        \vspace{4mm}
    \item \textbf{Fourier-based losses.}
        This group augments the standard $\text{Lp}$ loss in the time domain with an additional $\text{Lp}$ loss computed on the Fourier representations of the predicted and ground-truth signals, restricted to the $150$--$250$ Hz frequency band:
        
        \begin{equation}\label{eq:fourier_loss}
            \mathcal{L}(x, y) = \text{Lp}(x, y) + w\,\text{Lp}(\mathcal{F}_{\mathrm{mask}}(x), \mathcal{F}_{\mathrm{mask}}(y)),
        \end{equation}
        
        where $\mathcal{F}_{\mathrm{mask}}(\cdot)$ denotes the band-limited Fourier transform obtained by applying a Fast Fourier Transform followed by a frequency mask that isolates the $150$--$250$ Hz range. This band was selected because PVALB spiking dynamics in our data are concentrated in this regime. The additional spectral term encourages the model to match both the pointwise voltage values in the time domain and the oscillatory structure of the signal within the selected band. In particular, it penalizes discrepancies in spike timing and rapid voltage fluctuations that may be under-emphasized by purely time-domain losses, thereby improving the representation of high-frequency spiking dynamics.

\end{itemize}

\begin{table}[t]
    \vspace{-2mm}
    \centering
    \caption{Performance of the overfitting studies, reported as relative $\text{L2}$ training error for \texttt{NOBLE} on voltage traces and selected electrophysiological features under different training objectives.
        Row colors summarize scale-normalized performance: errors are min--max normalized within each metric, and then the mean and maximum normalized errors are computed across metrics.
        \begingroup\setlength{\fboxsep}{1.5pt}\colorbox{StrongRow}{\strut Green}\endgroup{} indicates mean normalized error $\leq 0.18$ and maximum normalized error $\leq 0.30$; \begingroup\setlength{\fboxsep}{1.5pt}\colorbox{MediumRow}{\strut blue}\endgroup{} indicates intermediate performance; and \begingroup\setlength{\fboxsep}{1.5pt}\colorbox{WeakRow}{\strut red}\endgroup{} indicates mean normalized error $\geq 0.35$ or maximum normalized error $\geq 0.75$.}
    \vspace{1.5mm}
    \label{tab:loss_function_ablation}
    \renewcommand{\arraystretch}{1.08}
    \setlength{\tabcolsep}{5pt}
    \resizebox{\linewidth}{!}{%
        \begin{tabular}{@{}lccccccc@{}}
            \toprule
            \textbf{Loss Function}
            & \textbf{Voltage}
            & \shortstack{\textbf{inv ISI}}
            & \shortstack{\textbf{Mean AP}\\\textbf{Amplitude}}
            & \shortstack{\textbf{Steady-State}\\\textbf{Voltage}}
            & \shortstack{\textbf{Mean}\\\textbf{Frequency}}
            & \shortstack{\textbf{AP}\\\textbf{width}}
            & \shortstack{\textbf{Spike}\\\textbf{count}} \\
            \midrule
            \rowcolor{MediumRow}
            Rel $\text{L2}$           & 0.29\%          & 0.04\%          & 1.95\%          & 0.05\%          & 0.04\%          & 2.88\%          & 0.00\% \\
            \rowcolor{MediumRow}
            Abs $\text{L2}$           & 0.27\%          & 0.05\%          & 2.06\%          & 0.04\%          & \textbf{0.00}\% & 2.37\%          & 0.00\% \\
            \rowcolor{StrongRow}
            Rel $\text{L4}$           & 0.25\%          & 0.03\%          & 0.79\%          & 0.06\%          & 0.02\%          & 0.56\%          & 0.00\% \\
            \rowcolor{MediumRow}
            Abs $\text{L4}$           & 0.28\%          & 0.02\%          & 1.09\%          & 0.09\%          & 0.02\%          & 2.01\%          & 0.00\% \\
            \rowcolor{MediumRow}
            Rel $\text{L6}$           & 0.33\%          & 0.02\%          & 0.89\%          & 0.10\%          & 0.02\%          & 1.24\%          & 0.00\% \\
            \rowcolor{WeakRow}
            Abs $\text{L6}$           & 0.44\%          & 0.02\%          & 1.44\%          & 0.11\%          & 0.02\%          & 2.02\%          & 0.00\% \\
            \rowcolor{StrongRow}
            Rel $\text{L4}$ ref-std   & 0.22\%          & 0.01\%          & 0.68\%          & 0.04\%          & 0.02\%          & 0.96\%          & 0.00\% \\
            \rowcolor{StrongRow}
            Abs $\text{L4}$ ref-std   & \textbf{0.16\%} & 0.02\%          & \textbf{0.51\%} & \textbf{0.03\%} & \textbf{0.00\%} & \textbf{0.55\%} & 0.00\% \\
            \rowcolor{StrongRow}
            Rel $\text{L4}$ ref-norm  & 0.19\%          & 0.02\%          & 0.63\%          & 0.04\%          & 0.02\%          & 1.01\%          & 0.00\% \\
            \rowcolor{StrongRow}
            Abs $\text{L4}$ ref-norm  & 0.20\%          & 0.02\%          & 0.67\%          & 0.04\%          & 0.02\%          & 0.90\%          & 0.00\% \\
            \rowcolor{WeakRow}
            Rel $\text{L4}$ both-std  & 0.45\%          & 0.03\%          & 1.53\%          & 0.09\%          & 0.02\%          & 1.58\%          & 0.00\% \\
            \rowcolor{StrongRow}
            Abs $\text{L4}$ both-std  & 0.22\%          & \textbf{0.00\%} & 0.74\%          & 0.05\%          & 0.02\%          & 0.68\%          & 0.00\% \\
            \rowcolor{MediumRow}
            Rel $\text{L4}$ both-norm & 0.26\%          & 0.02\%          & 0.77\%          & 0.05\%          & 0.02\%          & 1.05\%          & 0.00\% \\
            \rowcolor{StrongRow}
            Abs $\text{L4}$ both-norm & 0.25\%          & 0.03\%          & 0.81\%          & 0.06\%          & 0.03\%          & 1.14\%          & 0.00\% \\
            \midrule
            \rowcolor{MediumRow}
            Sigmoid-Gated Rel $\text{L2}$ $(w = 1)$            & 0.26\% & 0.03\% & 1.25\% & 0.06\%          & 0.04\% & 1.86\% & 0.00\% \\
            \rowcolor{MediumRow}
            Sigmoid-Gated Abs $\text{L2}$ $(w = 1)$            & 0.24\% & 0.03\% & 1.22\% & 0.05\%          & 0.01\% & 2.20\% & 0.00\% \\
            \rowcolor{MediumRow}
            Sigmoid-Gated Rel $\text{L4}$ $(w = 1)$            & 0.26\% & 0.02\% & 1.14\% & 0.06\%          & 0.02\% & 0.79\% & 0.00\% \\
            \rowcolor{MediumRow}
            Sigmoid-Gated Abs $\text{L4}$ $(w = 1)$            & 0.29\% & 0.02\% & 0.83\% & 0.07\%          & 0.02\% & 1.47\% & 0.00\% \\
            \rowcolor{StrongRow}
            Sigmoid-Gated Rel $\text{L2}$ $(w = 10)$           & 0.20\% & 0.02\% & 0.87\% & \textbf{0.03\%} & 0.02\% & 1.39\% & 0.00\% \\
            \rowcolor{MediumRow}
            Sigmoid-Gated Abs $\text{L2}$ $(w = 10)$           & 0.30\% & 0.03\% & 1.41\% & 0.05\%          & 0.02\% & 2.55\% & 0.00\% \\
            \rowcolor{MediumRow}
            Sigmoid-Gated Rel $\text{L4}$ $(w = 10)$           & 0.24\% & 0.03\% & 0.86\% & 0.07\%          & 0.02\% & 1.15\% & 0.00\% \\
            \rowcolor{MediumRow}
            Sigmoid-Gated Abs $\text{L4}$ $(w = 10)$           & 0.26\% & 0.02\% & 1.05\% & 0.07\%          & 0.03\% & 0.99\% & 0.00\% \\
            \rowcolor{WeakRow}
            Sigmoid-Gated Rel $\text{L2}$ $(w = 10, s = 0.01)$ & 0.48\% & 0.78\% & 0.91\% & 0.13\%          & 0.19\% & 0.88\% & 0.17\% \\
            \rowcolor{WeakRow}
            Sigmoid-Gated Abs $\text{L2}$ $(w = 10, s = 0.01)$ & 0.45\% & 0.02\% & 0.81\% & 0.11\%          & 0.02\% & 0.67\% & 0.00\% \\
            \rowcolor{MediumRow}
            Sigmoid-Gated Rel $\text{L2}$ $(w = 10, s = 0.1)$  & 0.23\% & 0.04\% & 1.33\% & 0.05\%          & 0.04\% & 2.07\% & 0.00\% \\
            \rowcolor{WeakRow}
            Sigmoid-Gated Abs $\text{L2}$ $(w = 10, s = 0.1)$  & 0.29\% & 0.22\% & 1.65\% & 0.05\%          & 0.29\% & 2.34\% & 0.26\% \\
            \midrule
            \rowcolor{MediumRow}
            Rel F2 $(w = 1)$   & 0.25\% & 0.06\%  & 1.74\% & 0.04\% & 0.07\% & 2.14\% & 0.08\% \\
            \rowcolor{WeakRow}
            Abs F2 $(w = 1)$   & 0.40\% & 0.12\%  & 3.45\% & 0.06\% & 0.14\% & 3.94\% & 0.09\% \\
            \rowcolor{MediumRow}
            Rel F4 $(w = 1)$   & 0.26\% & 0.02\%  & 0.98\% & 0.06\% & 0.02\% & 1.22\% & 0.00\% \\
            \rowcolor{WeakRow}
            Abs F4 $(w = 1)$   & 0.40\% & 0.03\%  & 1.41\% & 0.08\% & 0.02\% & 2.36\% & 0.00\% \\
            \rowcolor{MediumRow}
            Rel F2 $(w = 0.1)$ & 0.25\% & 0.06\%  & 1.72\% & 0.10\% & 0.07\% & 1.76\% & 0.08\% \\
            \rowcolor{WeakRow}
            Abs F2 $(w = 0.1)$ & 0.28\% & 19.13\% & 3.18\% & 0.05\% & 0.14\% & 3.45\% & 0.11\% \\
            \rowcolor{StrongRow}
            Rel F4 $(w = 0.1)$ & 0.25\% & 0.02\%  & 0.82\% & 0.06\% & 0.02\% & 1.03\% & 0.00\% \\
            \rowcolor{MediumRow}
            Abs F4 $(w = 0.1)$ & 0.38\% & 0.02\%  & 1.30\% & 0.08\% & 0.03\% & 2.03\% & 0.00\% \\
            \bottomrule
        \end{tabular}}
\end{table}

Several objectives fit the voltage traces reasonably well, but differ in how faithfully they preserve spike timing, spike amplitude, subthreshold dynamics, after-spike behavior, and non-spiking responses. Based on relative $\text{L2}$ training error and spike-related diagnostics, particularly action-potential amplitudes and widths, the reference-standardized absolute $\text{L4}$ loss and the sigmoid-gated relative $\text{L2}$ loss were among the candidates that performed best across voltage-level and feature-level metrics. The overfitting study serves only as an empirical guide, and the two strongest candidates were then trained at full scale (\Cref{sec:loss_results}). More broadly, \texttt{NOBLE} remains compatible with alternative loss functions when other electrophysiological properties are prioritized.

\begin{table}[h]
    \centering
    \caption{Test-set performance of the PVALB \texttt{NOBLE} model trained on two loss objectives selected following preliminary overfitting studies.}
    \vspace{1mm}
    \label{tab:full_loss_function_comparison}
    \renewcommand{\arraystretch}{1.08}
    \setlength{\tabcolsep}{5pt}
    \resizebox{\linewidth}{!}{%
        \begin{tabular}{@{}lccccccc@{}}
            \toprule
            \textbf{Loss function}
            & \textbf{Voltage}
            & \shortstack{\textbf{inv ISI}\\\textbf{values}}
            & \shortstack{\textbf{mean AP}\\\textbf{amplitude}}
            & \shortstack{\textbf{steady-state}\\\textbf{voltage}}
            & \shortstack{\textbf{AP}\\\textbf{width}}
            & \shortstack{\textbf{Spikecount}}
            & \shortstack{\textbf{Missed}\\\textbf{firing}}\\
            \midrule
            Abs \text{L4} ref-std & 2.63\% & 29.36\% & 15.71\% & 0.71\% & 13.11\% & 21.71\% & 3.11\% \\
            \midrule
            Sigmoid-Gated Rel \text{L2} (w = 10) & 1.73\% & 16.97\% & 8.73\% & 0.25\% & 10.54\% & 19.27\% & 1.87\% \\
            \bottomrule
        \end{tabular}}
    \vspace{4mm}
\end{table}

\subsection{Evaluation Metrics}

Although training minimizes the sigmoid-gated relative $\text{L2}$ loss, we report relative $\text{L2}$ error as a common and interpretable accuracy metric. Because small temporal shifts in spike timing can produce large waveform errors, we also evaluate physiologically meaningful summary features:

\begin{itemize}
    \item \texttt{inv\_ISI\_values}: inverse of the interspike interval values
    \item \texttt{mean\_AP\_amplitude}: mean amplitude of action potentials
    \item \texttt{steady\_state\_voltage}: average voltage following a stimulus
    \item \texttt{mean\_frequency}: average frequency of the firing rate
    \item \texttt{AP\_width}: width of spike at threshold
    \item \texttt{Spikecount}: total number of spikes in voltage trace
    \item missed firing: rate at which a predicted trace contains no spikes although its target fires
    \item missed non-firing: rate at which a predicted trace contains spikes although its target does not
\end{itemize}

For benchmarking, we compare \texttt{NOBLE} predictions against HoF model simulations and experimental recordings. HoF models provide the closest available biophysical approximations to experimental traces while preserving the biological variability needed for population-level evaluation. The machine-learning surrogates discussed in the main text are not designed to reproduce this HoF-induced variability and are therefore unsuitable baselines here. Because the PDE simulations used for training are themselves approximations of experimental recordings, reducing surrogate error far below the solver--experiment discrepancy risks fitting numerical artifacts rather than improving biological fidelity. We therefore tune hyperparameters to the solver--experiment error scale, beyond which further PDE-error reductions are unlikely to improve agreement with experimental data.

\clearpage

\section{Implementation Details and Scope}\label[appsec]{app:implementation}

We trained cell-type-specific \texttt{NOBLE} models based on a 1D Fourier Neural Operator (FNO)~\cite{li2020fno,duruisseaux2025FNOGuide}, implemented using the NeuralOperator library~\citep{kossaifi2024neural}. Each network has 6 layers, 96 hidden channels, and 256 Fourier modes, contributing to a total of 14.4M trainable parameters. Models were trained in PyTorch for 400 epochs using Adam with an initial learning rate of 0.002 and a ReduceLROnPlateau scheduler with decay factor 0.5 and patience of 540 steps. Each epoch comprised 95 batches, equivalent to 9,500 samples.
Each training and testing batch consisted of 100 samples.

\paragraph{Scope of the training design.}
We recognize that the present training design focuses specifically on inhibitory interneurons, DC step-current stimulation, and an FNO surrogate architecture. We retain the FNO used in the original \texttt{NOBLE} formulation to enable direct comparison with prior results, but the proposed conditioning and training framework is not tied to this backbone and could be paired with alternative neural operators or neural architectures. There is likewise no intrinsic reason to expect the framework to be better suited to inhibitory neurons than to excitatory neurons, for example. Furthermore, DC step currents are a standard protocol for characterizing neuronal responses and align naturally with the electrophysiological metrics used throughout this work, including \texttt{AP\_width}, \texttt{sag\_amplitude}, \texttt{time\_to\_first\_spike}, and the F-I relationship. Evaluating alternative surrogate backbones, excitatory neurons, and richer time-varying stimuli is left for future work.

\hfill \\

\section{Experimental Variability}\label[appsec]{app:experimental_variability}

\Cref{tab:cell_to_cell,tab:trial_to_trial} report the cell-to-cell and trial-to-trial variability within the experimental recordings, which \Cref{sec:test_performance} compares with the surrogate error in \Cref{tab:test_performance} and the PDE-to-experiment discrepancy in \Cref{tab:PDE_experiment}.

\begin{table}[h]
    \centering
    \caption{Relative $\mathrm{L2}$ cell-to-cell variability in experimental recordings across inhibitory neuron classes, evaluated using the same metrics as Table~\ref{tab:test_performance}.}
    \vspace{1mm}
    \label{tab:cell_to_cell}
    \renewcommand{\arraystretch}{1.08}
    \setlength{\tabcolsep}{5pt}
    \resizebox{\linewidth}{!}{%
        \begin{tabular}{@{}lccccccccc@{}}
            \toprule
            \textbf{Cell type}
            & \textbf{Voltage}
            & \shortstack{\textbf{inv ISI}\\\textbf{values}}
            & \shortstack{\textbf{mean AP}\\\textbf{amplitude}}
            & \shortstack{\textbf{steady-state}\\\textbf{voltage}}
            & \shortstack{\textbf{AP}\\\textbf{width}}
            & \shortstack{\textbf{Spikecount}}
            & \shortstack{\textbf{Missed}\\\textbf{firing}}
            & \shortstack{\textbf{Missed}\\\textbf{non-firing}}\\
            \midrule
            PVALB & 8.34\%  & 107.26\% & 7.69\%  & 8.40\%  & 18.50\% & 108.49\% & 9.23\%  & 15.18\% \\
            SST   & 10.54\% & 41.06\%  & 11.50\% & 11.25\% & 25.13\% & 117.43\% & 19.75\% & 33.33\% \\
            VIP   & 10.69\% & 70.46\%  & 13.22\% & 10.53\% & 24.56\% & 91.71\%  & 23.87\% & 19.14\% \\
            all   & 9.33\%  & 48.17\%  & 16.41\% & 9.30\%  & 32.07\% & 108.74\% & 19.44\% & 23.70\% \\
            \bottomrule
        \end{tabular}}
    \vspace{4mm}
\end{table}

\begin{table}[h]
    \centering
    \caption{Relative $\mathrm{L2}$ trial-to-trial variability in experimental recordings across inhibitory neuron classes, evaluated using the same metrics as Table~\ref{tab:test_performance}.}
    \vspace{1mm}
    \label{tab:trial_to_trial}
    \renewcommand{\arraystretch}{1.08}
    \setlength{\tabcolsep}{5pt}
    \resizebox{\linewidth}{!}{%
        \begin{tabular}{@{}lccccccccc@{}}
            \toprule
            \textbf{Cell type}
            & \textbf{Voltage}
            & \shortstack{\textbf{inv ISI}\\\textbf{values}}
            & \shortstack{\textbf{mean AP}\\\textbf{amplitude}}
            & \shortstack{\textbf{steady-state}\\\textbf{voltage}}
            & \shortstack{\textbf{AP}\\\textbf{width}}
            & \shortstack{\textbf{Spikecount}}
            & \shortstack{\textbf{Missed}\\\textbf{firing}}
            & \shortstack{\textbf{Missed}\\\textbf{non-firing}}\\
            \midrule
            PVALB & 4.43\% & 23.16\% & 4.25\% & 4.56\% & 6.07\% & 46.62\% & 20.00\% & 0.00\% \\
            SST   & 0.62\% & 6.95\%  & 3.08\% & 0.81\% & 2.35\% & 38.10\% & 0.00\%  & 33.33\% \\
            VIP   & 1.32\% & 17.89\% & 7.79\% & 1.56\% & 4.69\% & 10.56\% & 6.67\%  & 4.44\% \\
            \bottomrule
        \end{tabular}}
    \vspace{4mm}
\end{table}

\clearpage

\section{Spike Timing and Voltage Trace Examples}\label[appsec]{app:traces}

\begin{figure}[h] \vspace{12mm}
    \centering
    \includegraphics[width=0.49\linewidth]{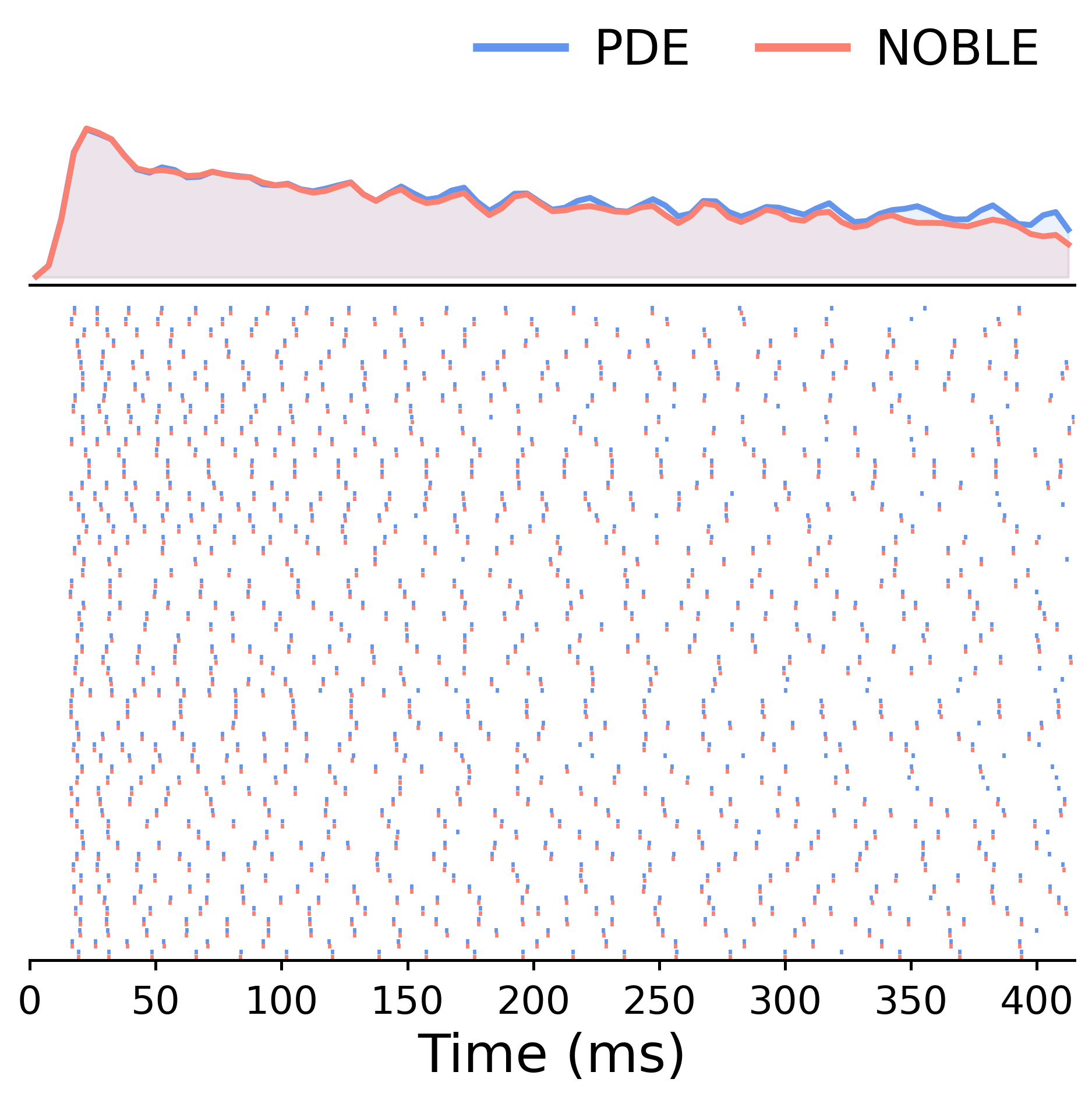}\hfill
    \includegraphics[width=0.49\linewidth]{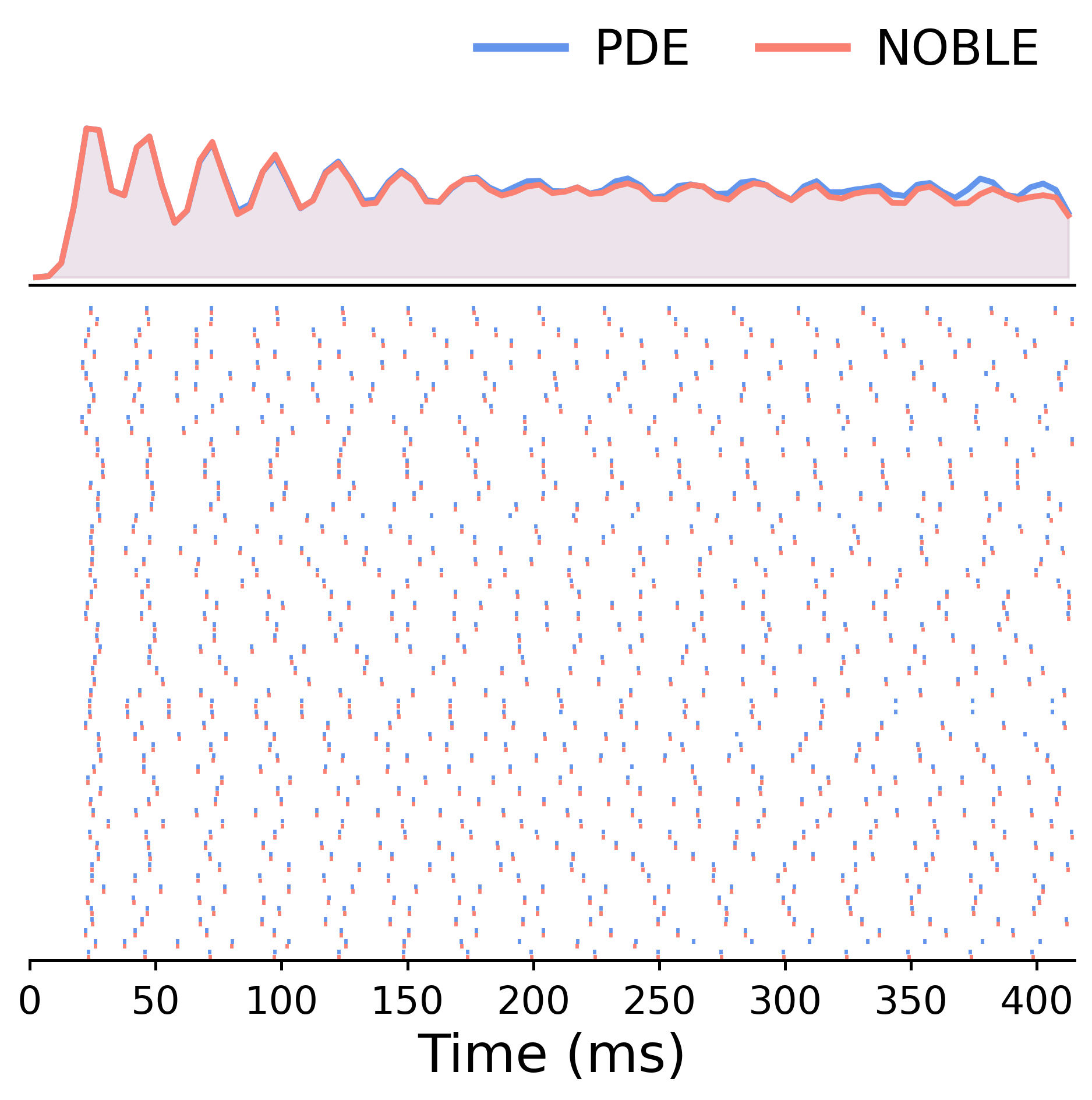}
    \caption{\textbf{Spike timing in SST and VIP neurons.} Raster comparison of spike times for SST (left) and VIP (right) neurons.
        Top panels show the smoothed population spike-time density for HoF simulations (blue) and \texttt{NOBLE} predictions (red).
        Bottom panels show individual spike events across traces.}
    \label{fig:rasters_sst_vip}
\end{figure}

\Cref{fig:rasters_sst_vip} complements the PVALB rasters of \Cref{fig:rasters} with raster comparisons for SST and VIP neurons. As for PVALB, the \texttt{NOBLE} spike times closely follow the HoF simulations at both the single-trace and population levels. 

\hfill 

\begin{figure}[h]
    \centering
    \includegraphics[width=\linewidth]{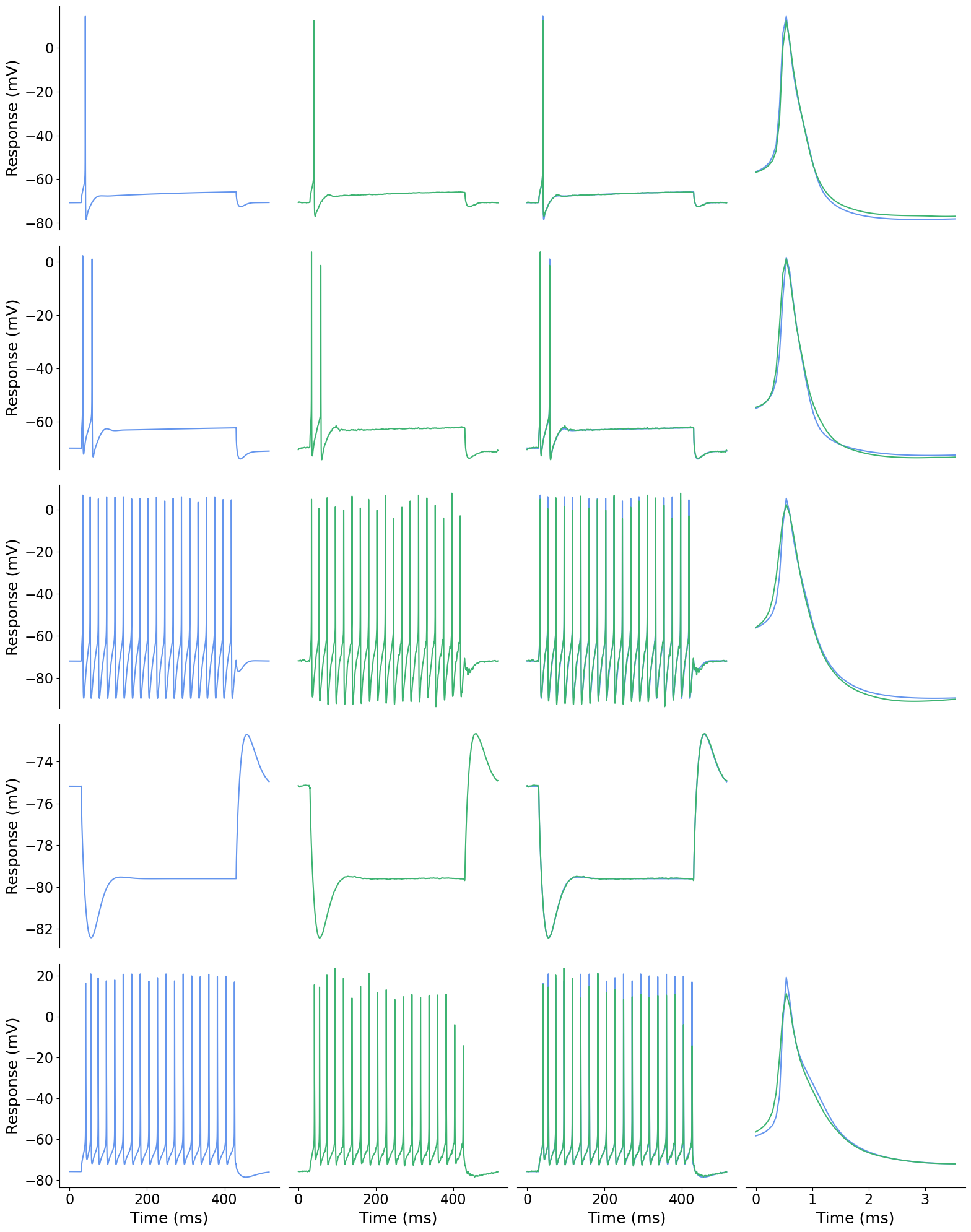}
    \caption{From left to right, columns show: (1) PDE-simulated voltage traces; (2) \texttt{NOBLE} voltage trace predictions; (3) overlay of PDE and \texttt{NOBLE} predictions; and (4) spike-triggered voltage averages from PDE simulations and \texttt{NOBLE}.
        Voltage traces illustrate high spike-timing agreement between \texttt{NOBLE} predictions and PDE simulations.}
    \label{fig:spike-timing-examples}
\end{figure}

\Cref{fig:spike-timing-examples} shows representative voltage traces comparing \texttt{NOBLE} predictions with PDE-simulated responses. Across test set examples, \texttt{NOBLE} closely aligns with the timing of individual spikes while also preserving the overall subthreshold voltage trajectory. These traces provide qualitative evidence that the surrogate captures spike timing, not only coarse firing-rate behavior.

\hfill \\

\vspace{5mm}

\clearpage

\section{Held-Out HoF Trace Examples}\label[appsec]{app:hoftest}

\begin{figure}[h]
    \centering
    \includegraphics[width=\linewidth]{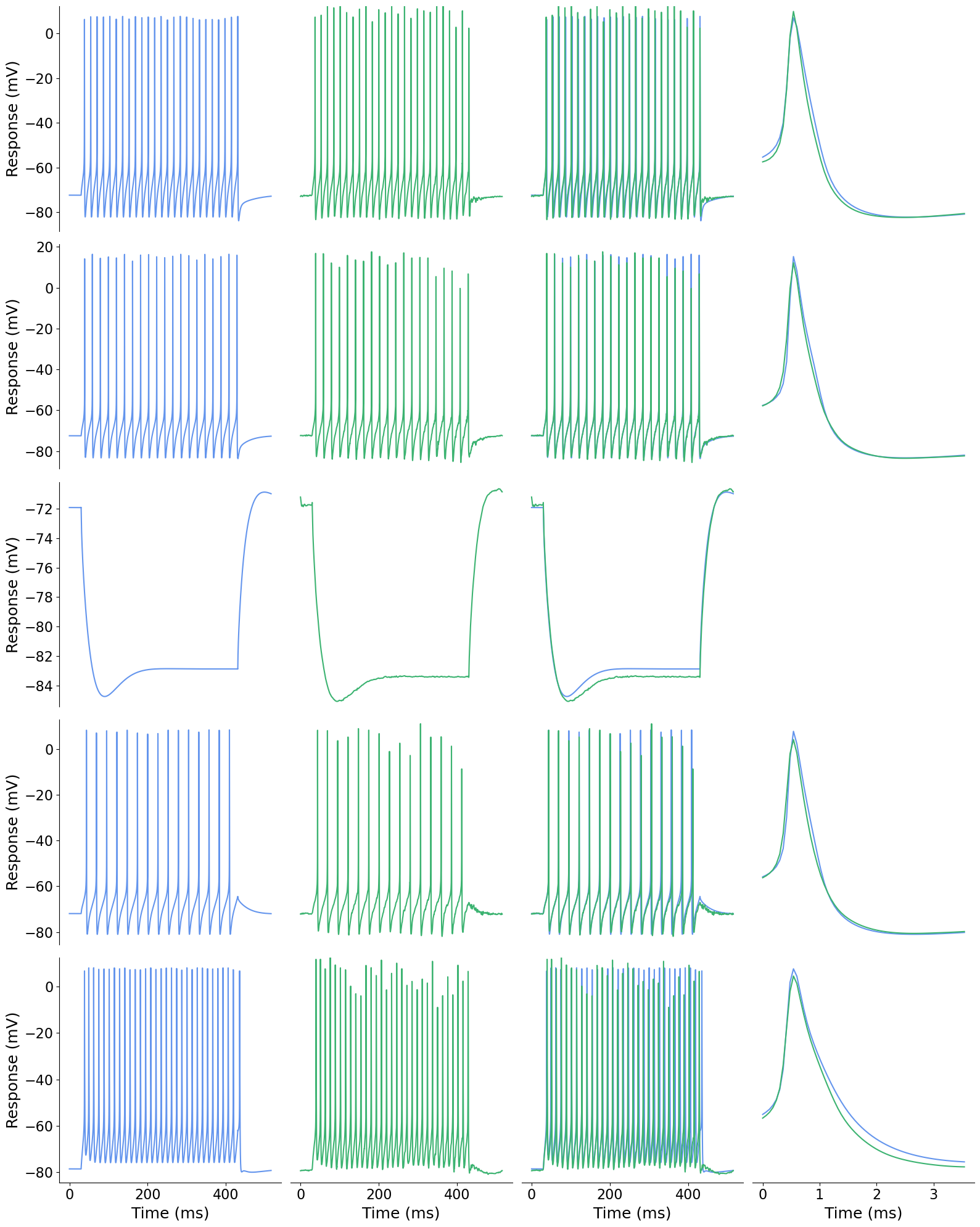}
    \caption{\texttt{NOBLE} predictions on held-out HoF models of cells represented during training maintain qualitative interspike timing and general waveform dynamics.}
    \label{fig:out_of_distribution_examples}
\end{figure}

\Cref{fig:out_of_distribution_examples} shows example traces from held-out HoF models of cells represented during training, comparing \texttt{NOBLE} predictions with PDE-simulated responses. \texttt{NOBLE} preserves key features, including interspike timing as shown in Table~\ref{tab:out_of_distribution}, although minor perturbations in the predicted traces can accumulate across subsequent action potentials. Despite these accumulated deviations, the individual voltage traces exhibit high levels of qualitative agreement with the PDE-based targets.

\vspace{8mm}

\end{document}